\documentclass[superscriptaddress,groupedaddress,nofootnoteinbib,11pt]{article}
\pdfoutput=1 
\usepackage{jheppub}

\usepackage[utf8x]{inputenc}
\usepackage{mathtools,slashed,mathrsfs}
\usepackage[caption=false]{subfig}
\usepackage{dcolumn}
\usepackage{multirow}
\usepackage{tabularx}
\usepackage{booktabs}
\usepackage{bm}
\usepackage{comment}
\usepackage{setspace}
\usepackage[dvipsnames]{xcolor}
\usepackage[normalem]{ulem} 
\usepackage{enumerate}
\usepackage{siunitx}
\usepackage{xspace}
\usepackage{adjustbox}
\allowdisplaybreaks

\newcommand{\Sec}[1]{Sec.~\ref{#1}}

\newcommand{\App}[1]{Appendix~\ref{#1}}

\newcommand{\Eq}[1]{Eq.~(\ref{#1})}
\newcommand{\Eqs}[2]{Eqs.~(\ref{#1}) and (\ref{#2})}
\newcommand{\Eqst}[2]{Eqs.~(\ref{#1})--(\ref{#2})}

\newcommand{\beq}{\begin{equation}}
\newcommand{\eeq}{\end{equation}}
\newcommand{\ba}{\begin{array}}
\newcommand{\ea}{\end{array}}
\newcommand{\bea}{\begin{eqnarray}}
\newcommand{\eea}{\end{eqnarray} }
\newcommand{\be}{\begin{eqnarray}}
\newcommand{\ee}{\end{eqnarray}}
\newcommand{\bal}{\begin{align}}
\newcommand{\eal}{\end{align}}
\newcommand{\bi}{\begin{itemize}}
\newcommand{\ei}{\end{itemize}}
\newcommand{\ben}{\begin{enumerate}}
\newcommand{\een}{\end{enumerate}}
\newcommand{\bc}{\begin{center}}
\newcommand{\ec}{\end{center}}
\newcommand{\bt}{\begin{table}}
\newcommand{\et}{\end{table}}
\newcommand{\btb}{\begin{tabular}}
\newcommand{\etb}{\end{tabular}}

\newcommand{\bl}{\left}
\newcommand{\br}{\right}

\newcommand{\ie}{\textit{i.e.}\ }

\newcommand{\eV}{\mathrm{eV}}
\newcommand{\keV}{\mathrm{keV}}
\newcommand{\MeV}{\mathrm{MeV}}
\newcommand{\GeV}{\mathrm{GeV}}
\newcommand{\TeV}{\mathrm{TeV}}

\newcommand{\Mpc}{\mathrm{Mpc}}

\newcommand{\km}{\mathrm{km}}
\newcommand{\seg}{\mathrm{s}}

\newcommand{\mO}{\mathcal{O}}

\newcommand{\Lag}{\mathcal{L}}
\newcommand{\mH}{\mathcal{H}}
\newcommand{\mM}{\mathcal{M}}
\newcommand{\mS}{\mathcal{S}}
\newcommand{\mD}{\mathcal{D}}
\newcommand{\mDH}{\mathcal{DH}}

\newcommand{\mDHS}{\mathcal{DHS}}

\newcommand{\vpi}{\ensuremath{\varpi}}

\newcommand{\eq}{\ensuremath{\mathrm{eq}}}

\newcommand{\tr}{\mathrm{tr}}

\newcommand{\lcdm}{\ensuremath{\Lambda\mathrm{CDM}}\xspace}
\newcommand{\spa}{SPartAcous\xspace}

\newcommand{\spap}{SPartAcous+\xspace}

\newcommand{\spapt}{SPartAcous+3\xspace}

\newcommand{\dm}{\mathrm{dm}}
\newcommand{\cdm}{\mathrm{CDM}}
\newcommand{\idm}{\mathrm{idm}}
\newcommand{\dr}{\mathrm{dr}}

\newcommand{\Neff}{\ensuremath{N_{\mathrm{eff}}}\xspace}
\newcommand{\DNeff}{\ensuremath{\Delta N_\mathrm{eff}}\xspace}
\newcommand{\IR}{\ensuremath{\mathrm{IR}}}
\newcommand{\UV}{\ensuremath{\mathrm{UV}}}
\newcommand{\NIR}{\ensuremath{\DNeff^\text{IR}}\xspace}
\newcommand{\NUV}{\ensuremath{\DNeff^\text{UV}}\xspace}
\newcommand{\HO}{\ensuremath{H_0}\xspace}
\newcommand{\Se}{\ensuremath{S_8}\xspace}

\newcommand{\tc}{\ensuremath{\tau_c}\xspace}
\newcommand{\dtc}{\ensuremath{\dot\tau_c}\xspace}
\newcommand{\Trx}{\ensuremath{\Theta_{\dr\text{--}\idm}}\xspace}
\newcommand{\dTrx}{\ensuremath{\dot\Theta_{\dr\text{--}\idm}}\xspace}

\begin{document}
\preprint{\begin{flushright}
    UTWI-19-2023\\
\end{flushright}}

\title{Stepped Partially Acoustic Dark Matter: Likelihood Analysis and Cosmological Tensions}

\date{\today}
\author[a,c]{Manuel A. Buen-Abad,}
\author[a]{Zackaria Chacko,}
\author[b]{Can Kilic,}
\author[a, d]{Gustavo Marques-Tavares,}
\author[b]{Taewook Youn}

\affiliation[a]{Maryland Center for Fundamental Physics, Department of Physics, University of Maryland, College Park, MD 20742, U.S.A.}
\affiliation[b]{Center for Theory, Weinberg Institute, Department of Physics, University of Texas at Austin, Austin, TX 78712, U.S.A.}
\affiliation[c]{Dual CP Institute of High Energy Physics, C.P. 28045, Colima, M\'{e}xico}
\affiliation[d]{Department of Physics and Astronomy, University of Utah, Salt Lake City, UT 84112, U.S.A.}
\emailAdd{buenabad@umd.edu}
\emailAdd{zchacko@umd.edu}
\emailAdd{kilic@physics.utexas.edu}
\emailAdd{g.marques@utah.edu}
\emailAdd{taewook.youn@utexas.edu}

\abstract{
We generalize the recently proposed Stepped Partially Acoustic Dark Matter (SPartAcous) model by including additional massless degrees of freedom in the dark radiation sector. We fit SPartAcous and its generalization against cosmological precision data from the cosmic microwave background, baryon acoustic oscillations, large-scale structure, supernovae type Ia, and Cepheid variables. We find that SPartAcous significantly reduces the $H_0$ tension but does not provide any meaningful improvement of the $S_8$ tension, while the generalized model succeeds in addressing both tensions, and provides a better fit than \lcdm and other dark sector models proposed to address the same tensions. In the generalized model, $H_0$ can be raised to $71.4 ~ \km/\seg/\Mpc$ (the 95\% upper limit), reducing the tension, if the fitted data does not include the direct measurement from the SH0ES collaboration, and to $73.7~\km/\seg/\Mpc$ (95\% upper limit) if it does. A version of {\tt CLASS} that has been modified to analyze this model is publicly available at \href{https://github.com/ManuelBuenAbad/class\_spartacous}{\tt github.com/ManuelBuenAbad/class\_spartacous}.
}

\maketitle


\section{Introduction}
\label{sec:intro}

Over the last decade, cosmology has entered a golden age in which cosmological observables have been measured to an unprecedented level of precision. While $\lcdm$, the standard model of cosmology, has been able to successfully describe a broad range of measurements over this period, in recent years there has been increasing tension between some of the most precise experimental results. The greatest source of this tension arises from the differences in the various measurements of $H_0$, the current expansion rate of the universe. Indirect measurements involving a fit of $\lcdm$ to the cosmic microwave background (CMB)~\cite{Planck:2018vyg, ACT:2020gnv, SPT-3G:2021eoc}
and large scale structure (LSS) data~\cite{DAmico:2019fhj,Ivanov:2019pdj,Philcox:2020vvt} favor lower values of $H_0 \lesssim 68$ km/s/Mpc~\cite{Planck:2018vyg, ACT:2020gnv, SPT-3G:2021eoc, DAmico:2019fhj,Ivanov:2019pdj,Philcox:2020vvt,Schoneberg:2019wmt, Addison:2013haa, Aubourg:2014yra, Addison:2017fdm, Blomqvist:2019rah, Cuceu:2019for, Verde:2016ccp, Bernal:2021yli}
than more direct methods based on the so-called cosmic ladder of standard candles such as Type Ia Supernovae (SN)~\cite{Riess:2021jrx, Freedman:2019jwv, Freedman:2021ahq, Yuan:2019npk, Soltis:2020gpl, Khetan:2020hmh, Huang:2019yhh, LIGOScientific:2019zcs, Schombert:2020pxm, Anderson:2023aga, Pesce:2020xfe,Dainotti:2023bwq}, which prefer $H_0 \gtrsim 70$ km/s/Mpc~(see also Ref.~\cite{Wong:2019kwg} for $H_0$ measurements using strongly lensed quasars which also find large values for $H_0$ in tension with CMB, although there are potential systematics that could be impacting these measurements~\cite{Birrer:2020tax}). 
Comparing the most precise measurements in each of these two categories, namely the \lcdm~fit to Planck CMB data~\cite{Planck:2018vyg} which gives
$H_0 = 67.36 \pm 0.54$ km/s/Mpc,
and the supernovae measurements made by the SH0ES collaboration calibrated to Cepheid variable stars~\cite{Riess:2021jrx} which yield
$H_0 = 73.04 \pm 1.04$ km/s/Mpc, the tension has reached $5\sigma$ significance~\cite{Verde:2019ivm, DiValentino:2021izs, Schoneberg:2021qvd}.
Another long standing source of tension, albeit more modest in significance than that of $H_0$, involves the amplitude of the matter power spectrum at relatively small scales, conventionally expressed in terms of the $S_8$ parameter. Direct measurements of this parameter, which is defined in terms of the matter energy density fraction $\Omega_m$ and the variance $\sigma_8^2$ of matter overdensities at $8~\Mpc/h$ as $S_8 \equiv \sigma_8 \sqrt{\Omega_m/0,3}$, have also consistently been in $2-3 \, \sigma$ tension with the value inferred from the $\lcdm$ fit to the CMB.\footnote{Note that a recent joint analysis by the DES and KiDS-1000 collaborations finds a smaller (1.7 $\sigma$) tension, even though their separate analysis find a more significant tension~\cite{DES:2021wwk, Li:2023azi}.} These tensions could be the first evidence of the need to depart from the \lcdm paradigm, and have motivated a considerable effort in the study of extensions of the standard model of cosmology that can accommodate them. For a sample of proposals that aim to solve the $H_0$ tension, see Refs.~\cite{Zhao:2017cud,DiValentino:2017gzb,Poulin:2018cxd,Smith:2019ihp,Lin:2019qug,Alexander:2019rsc,Agrawal:2019lmo,Escudero:2019gvw,Berghaus:2019cls,Ye:2020btb,RoyChoudhury:2020dmd,Brinckmann:2020bcn,Krishnan:2020vaf,Das:2020xke,Niedermann:2021vgd, Aloni:2021eaq, Dainotti:2021pqg,Odintsov:2022eqm,Berghaus:2022cwf,Colgain:2022rxy, Brinckmann:2022ajr,Sandner:2023ptm}. A more comprehensive list may be found in Refs.~\cite{DiValentino:2021izs, Schoneberg:2021qvd, Abdalla:2022yfr,Poulin:2023lkg} and references therein. Proposals to solve the $S_8$ tension include, for example, Refs.~\cite{Battye:2014qga,Buen-Abad:2015ova,Lesgourgues:2015wza,Murgia:2016ccp,Kumar:2016zpg,Chacko:2016kgg,Buen-Abad:2017gxg,Buen-Abad:2018mas,Dessert:2018khu,Archidiacono:2019wdp,Heimersheim:2020aoc,Bansal:2021dfh,Enqvist:2015ara,Poulin:2016nat,Clark:2020miy,FrancoAbellan:2021sxk}.
Unfortunately, many of the most promising proposals to address the $H_0$ tension lead to an increase in $S_8$, making the latter tension more significant. It is therefore important to consider models that can simultaneously address both tensions. Efforts in this direction include Refs.~\cite{Ye:2021iwa,Schoneberg:2022grr,Joseph:2022jsf,Buen-Abad:2022kgf,Wang:2022nap,Bansal:2022qbi,Zu:2023rmc,Cruz:2023lmn}.

In a recent publication we put forward a new joint solution to the \HO and \Se tensions, the {\bf S}tepped {\bf Part}ially {\bf Acous}tic Dark Matter model, (``SPartAcous'')~\cite{Buen-Abad:2022kgf}. This scenario naturally combines two mechanisms that have been proposed to alleviate the cosmological tensions in interacting dark sector models, namely a dark radiation (DR) bath with a mass-threshold~\cite{Aloni:2021eaq} that leads to a step-like feature in the fractional energy density in radiation, and a subcomponent of dark matter which is kinetically coupled to this DR~\cite{Chacko:2016kgg,Buen-Abad:2017gxg}. The change in the energy density in radiation can address the Hubble tension by decreasing the sound horizon \cite{Bernal:2016gxb,Aylor:2018drw,Knox:2019rjx}. At the same time the interactions between dark matter and DR give rise to Dark Acoustic Oscillations (DAOs) that suppress structure at small scales, and can thereby help resolve the $S_8$ tension \cite{Buen-Abad:2022kgf}. 
The presence of the mass threshold distinguishes \spa from {\bf P}artially {\bf Ac}oustic {\bf D}ark {\bf M}atter (``PAcDM")~\cite{Chacko:2016kgg,Buen-Abad:2017gxg}, and leads to very different predictions for the matter power spectrum. The reason for the difference is that the interacting dark matter subcomponent decouples from the DR once the temperature falls below the mass threshold, so that while there is little deviation from \lcdm on larger length scales, the amplitude at the scales relevant to \Se is reduced. This latter feature distinguishes \spa from most other models that address the Hubble tension by increasing the energy density in radiation, since they typically enhance the matter power spectrum at scales relevant for $\Se$, worsening the $S_8$ tension.

In this work we quantitatively investigate how well \spa fits a wide range of cosmological data compared to \lcdm, and to what extent it can solve the $H_0$ and \Se tensions. We also propose a simple generalization of the model, labelled \spap, in which we enlarge the number of massless states in the dark sector, thereby altering the size of the step. As with \spa, we evaluate how well \spap fits the data and addresses the tensions. We have implemented these models in a modified version of the Boltzmann code {\tt CLASS}, which we make publicly available at \href{https://github.com/ManuelBuenAbad/class\_spartacous}{\tt github.com/ManuelBuenAbad/class\_spartacous}. Our implementation includes a generalization of the tight-coupling approximation~\cite{Peebles:1970ag,Ma:1995ey,Blas:2011rf} for a fluid that has a mass threshold, which speeds up the code. This is described in detail in \App{sec:appappx}. Our implementation also includes a correction to the superhorizon initial conditions of the adiabatic perturbations due to the time-dependence of the equation of state, which was missed in earlier work on dark sector models with a mass threshold.

This paper is organized as follows. In \Sec{sec:model} we review the original \spa model and present its generalization, \spap. \Sec{sec:results} sets forth the implementation of the \spa and \spap models in {\tt CLASS}, lists the data we used to fit the models, and describes the results. Our conclusions are summarized in \Sec{sec:concl}. Details about our {\tt CLASS} code are expounded upon in \App{sec:apprecap} (equations and initial conditions), and \App{sec:appappx} (approximations), while further numerical results and tables are included in \App{sec:appnum}.

\section{The Model}
\label{sec:model}

In this section, we first quickly review the \spa model proposed in Ref.~\cite{Buen-Abad:2022kgf} and then present a simple generalization, \spap. 
The original \spa model adds three additional fields to \lcdm: a new massless (Abelian) gauge field $A$, and two new fields charged under this gauge group, a light vector-like fermion $\psi$ and a heavy scalar $\chi$. The heavy scalar $\chi$ constitutes a subcomponent of dark matter, which we will label interacting dark matter (iDM) due to its interactions with DR. The primary component of dark matter is assumed to be standard collisionless cold dark matter (CDM) (see Ref.~\cite{Chacko:2016kgg} for a way to implement both CDM and iDM components within a single theoretical framework). We take the fermion $\psi$ to be sufficiently light that it behaves as DR at early times and only becomes non-relativistic during the CMB epoch. Once the dark sector temperature drops below the mass of $\psi$, it annihilates away into gauge bosons. Due to its entropy being transferred to the remaining radiation, we obtain a step-like increase in the energy density in DR at that time. As is conventional, we parametrize the energy density in radiation in units of \Neff, the effective number of neutrinos. 
The part of the Lagrangian describing the model is given by,
\beq
\label{eq:lagrangian}
    \Lag_\text{dark} = -\frac{1}{4} V_{\mu\nu}V^{\mu\nu} + \bar \psi (i \slash  \!\!\!\! D - m_\psi) \psi + |D \chi|^2 - m_\chi^2 |\chi|^2\, ,
\eeq
where $V_{\mu \nu}$ is the field strength associated with the new gauge boson $A$, and $D_\mu \equiv \partial_\mu + i g_d A_\mu$ is the covariant derivative (with $g_d$ being the associated gauge coupling).

The dark sector is decoupled from the Standard Model plasma (at the times of interest), and therefore has its own temperature $T_d$. This temperature is directly related to the contribution of the DR to $\Delta \Neff$ at late times, which we denote by $\NIR$. The scales we are most concerned with are the ones that enter the horizon not much earlier than matter-radiation equality, at which time $\chi$ was already non-relativistic with a cosmic abundance set at a much earlier time. We will express the abundance of $\chi$ in terms of its fractional contribution to the total energy density in dark matter,
\begin{equation*}
    f_\chi = \frac{\rho_\chi}{\rho_\cdm + \rho_\chi} \, ,
\end{equation*}
where $\rho_\cdm$ corresponds to the energy density in the cold (non-interacting) dark matter component (CDM).

\begin{figure}[tb]
 	\centering
 	\includegraphics[width=.4\linewidth]{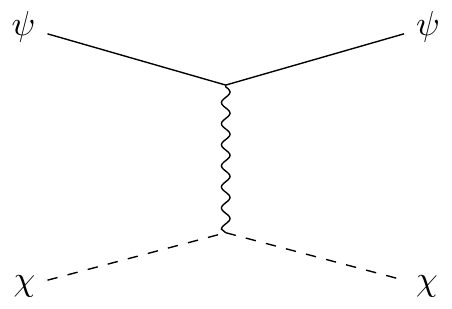}
 	\includegraphics[width=.4\linewidth]{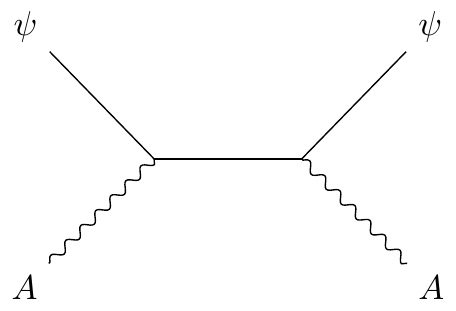}
 	\caption{Feynman diagrams for the most relevant interactions in the dark sector fluid. The left diagram is responsible for keeping the interacting dark matter component $\chi$ in equilibrium with the DR. The diagram on the right is responsible for keeping $\psi$ in equilibrium with the gauge boson $A$.}
 	\label{fig:diagrams}
 \end{figure}

At temperatures above $m_\psi$, the DR and iDM components behave as a single fluid due to the tight coupling between $\psi$ and $\chi$ mediated by the gauge interactions, shown in the first Feynman diagram of Fig.~\ref{fig:diagrams}, and due to Compton scattering between $\psi$ and $A$, shown in the second diagram. We will take $m_\chi$ to be sufficiently heavy that Compton scattering between $A$ and $\chi$ is not efficient at the temperatures relevant for the CMB. This ensures that once the dark sector's temperature falls below $m_\psi$, the dynamics changes in two significant ways. Firstly, the energy density in DR increases due to the annihilation of $\psi$, creating a step in \Neff, in a manner analogous to that of the Wess Zumino Dark Radiation (``WZDR'') model~\cite{Aloni:2021eaq}. Secondly, the interactions between DR and iDM decouple due to the exponential suppression in the number density of $\psi$
so that, from this point onward, the iDM becomes collisionless and evolves identically to CDM. This behavior stands in sharp contrast to that of WZDR+, a generalization of WZDR first proposed in Ref.~\cite{Joseph:2022jsf}, in which all of the dark matter interacts very weakly with the DR, eventually decoupling from it in a much slower manner. The redshift $z_t$ at which $T_d = m_\psi$ characterizes the time of the transition, and it will be used in our parameter scans in place of $m_\psi$. The dark gauge coupling $g_d$ is, in principle, another important parameter of the model. However, as we will discuss in~\Sec{sec:results}, because the couplings considered are sufficiently large as to make the DR and iDM behave as a single tightly-coupled fluid, the precise value of the coupling is largely irrelevant and it only plays a role in determining exactly when $\psi$ freezes out and $\chi$ decouples from the DR. Since both of those changes occur due to the number density of $\psi$ becoming exponentially suppressed, the dependence of the decoupling temperature on $g_d$ is only logarithmic, leading to minimal dependence of the physics on the value of $g_d$. Even after $\psi$ has frozen out, the DR is prevented from free streaming by a self-interaction among the gauge bosons of the Euler-Heisenberg form, which is generated at loop level when $\psi$ is integrated out \cite{Buen-Abad:2022kgf}.

\subsection{\spap}
\label{sec:spap}

A simple way to generalize the \spa model is to allow for different values in the size of the step in \DNeff at the mass threshold, while retaining the decoupling of DR and iDM. The step size, i.e. the fractional change in \DNeff, is related to the fractional change in the number of relativistic degrees of freedom, $r_g$, as
\beq
    \label{eq:step-size}
    r_g \equiv \bl( \frac{g^*_\UV - g^*_\IR}{g^*_\IR} \br) = \bl(\frac{\NIR}{\NUV}\br)^3 - 1 \ ,
\eeq
where the superscript $\UV$ $(\IR)$ corresponds to quantities above (below) the mass threshold, and $g^*$ represents the effective number of relativistic degrees of freedom in the dark sector interacting fluid. For the original \spa model, $g^*_\UV = 11/2$ and $g^*_\IR = 2$, and so \DNeff increases by approximately $40\%$ at the step. As will be shown in the next section, with such a large step, the data does not allow for a non-negligible fraction of the dark matter energy density to be iDM, and the best fit approaches a WZDR limit in which there is effectively no iDM. This motivates us to consider a generalization of our original model with a reduced step size. A decrease in the size of the step can be realized by increasing the number of massless particles in the interacting dark fluid.

In generalizing the \spa model, we will introduce an additional Abelian gauge field $A^\prime$ and $N_{\rm df}$ new vector-like massless fermions $\xi_i$, charged under $A^\prime$, which together constitute an extra DR component. $\psi$ is also charged under this new $U(1)$ gauge symmetry. On the other hand, $\chi$ only couples to $A$, and the $\xi_i$ only couple to $A'$.\footnote{While the model allows for a broad range of choices for the charge assignments of the $\chi$, $\psi$, and $\xi_i$ particles under the two $U(1)$'s, the cosmology only depends on when the relevant processes go out of equilibrium. For simplicity, we choose all charges to be $\pm 1$.} This ensures that after the mass threshold, the interacting dark matter component decouples from the dark radiation. The new terms added to the Lagrangian are therefore:
\beq
\label{eq:newlagrangian}
    \Lag_\text{dark}^\prime = -\frac{1}{4} V^\prime_{\mu\nu}V^{\mu\nu\prime} +  \sum_i^{N_{\rm df}} i \bar \xi_i \gamma^\mu (\partial_\mu + i g_d^\prime A^\prime_\mu) \xi_i \, , 
\eeq
where $V_{\mu \nu}^\prime$ is the field strength associated with the additional gauge boson $A^\prime$, whose gauge coupling is $g_d^\prime$. In addition, the covariant derivative acting on $\psi$ in \Eq{eq:lagrangian} is modified to include $A^\prime$. We only require that $\alpha_d^\prime$ be large enough for the new massless particles $\xi$ to be in equilibrium with $A'$ at the time when the smallest length scales of interest enter the horizon. This condition is easily satisfied over a broad range for $\alpha_d^{\prime}$ \cite{Buen-Abad:2022kgf}:
\beq\label{eq:alphad}
    \alpha_d^{\prime 2} T_d \gtrsim \frac{T^2}{M_p} \quad \Rightarrow \quad \alpha_d^\prime \gtrsim 10^{-12} \, \bl( \frac{T}{1~\keV} \br)^{1/2} \ ,
\eeq
where we have used the fact that $T_d$ is not significantly different from $T$.

Together the gauge interactions mediated by $A$ and $A^\prime$ ensure that the entire dark sector behaves like a tightly coupled fluid down until the mass threshold. This alters the values of $g^*_{\UV/\IR}$ in Eq.~(\ref{eq:step-size}). Because we are adding vector-like fermions, $g^*_{\UV/\IR}$ change by multiples of $7/2$, and therefore the fractional change in the number of degrees of freedom as a function of $N_{\rm df}$ is given by
\beq
    \label{eq:step-flavor}
    r_g = \bl(\frac{7}{8+7 \, N_{\rm df}} \br) \ .
\eeq
At the same time, because the new fermions $\xi_i$ are not charged under $A$, they do not scatter efficiently with iDM. This ensures that the model retains the important feature that the interactions between iDM and DR decouple shortly below the mass threshold. 

The presence of $\psi$, a state charged under both gauge groups, will nevertheless lead to a kinetic mixing between $A$ and $A'$ of size $\epsilon \sim g_d g_d^\prime / (16 \pi^2)$, thereby inducing a coupling between the new massless fermions and the iDM at loop level. However, in the range of $\alpha_d$ and $\alpha_d^{\prime}$ of interest, this interaction is too small to keep the iDM in equilibrium with the DR, and can be safely neglected. To be more quantitative, comparing the interaction rate $\dot{Q}_{ \xi \chi}$ to $H$ leads to the criterion
\bea
 \label{eq:Qchipsi}
    \dot{Q}_{\xi \chi} & \sim &  N_{\rm df} \frac{\alpha_d^2 \alpha_d^{\prime2} T_d^2}{16 \pi^2 m_\chi} \  \lesssim \   H \sim \frac{T^2}{M_p} \nonumber\\
    \Rightarrow \alpha_d^{\prime} & \lesssim & 10^{-4} \, \bl( \frac{10^{-3}}{\alpha_d}\br) \bl( \frac{m_\chi}{10^3~\GeV} \br)^{1/2} \bl( \frac{3}{N_{\rm df}} \br)^{1/2} \ .
\eea
\Eqs{eq:alphad}{eq:Qchipsi} show that as long as $10^{-12} < \alpha_d^\prime < 10^{-4}$ (for these benchmark values of $\alpha_d$ and $m_\chi$), the DR subcomponent consisting of $A'$ and $\xi_i$ will remain self-interacting but decouple from the iDM.\footnote{An alternative way to modify the \spa model to decrease the step size is to add a very weakly coupled, unconfined, $\mathrm{SU}(N_{\rm dc})$ dark color gauge field, with $\psi$ belonging to the fundamental representation. In this case, the dark gluons would play the role of the extra DR, coupled to $\psi$ via the gauge interaction. After the temperature drops below $m_\psi$, the dark gluons would still behave as an interacting fluid due to their self-interactions. We will not explore this possibility further in this paper; however, for the same numerical value of the step size, which is a function of $r_g = 7 / (4 N_{\rm dc})$ in this case, the cosmological history would be identical to the \spap model.}

\section{Results}
\label{sec:results}

We have modified the CMB code {\tt CLASS}~\cite{Lesgourgues:2011re,Blas:2011rf,Lesgourgues:2011rg,Lesgourgues:2011rh} to include the \spa and \spap models described in the last section. We have implemented a number of approximations, analogous to the ones described in~\cite{Blas:2011rf}, in order to sufficiently speed up the code to allow an efficient exploration of the parameter space. We describe these approximations in Appendix~\ref{sec:appappx}. Our code is publicly available at \href{https://github.com/ManuelBuenAbad/class\_spartacous}{\tt github.com/ManuelBuenAbad/class\_spartacous}. We employ this modified version of {\tt CLASS}, combined with the MCMC sampler {\tt MontePython}~\cite{Audren:2012wb,Brinckmann:2018cvx} to investigate how well the model fits different combinations of data sets and to find the allowed parameter regions for the associated cosmological parameters. We use the Metropolis-Hastings algorithm in {\tt MontePython}, and the resulting MCMC chains are considered to have converged if the Gelman-Rubin (GR) criterion $R < 1.01$ is satisfied, where $R$ is the GR statistic~\cite{Gelman:1992zz}.

In addition to the standard \lcdm parameters $\{ \omega_b, \, \omega_\dm, \, \theta_s, \, \ln \bl( 10^{10} A_s \br), \, n_s, \, \tau_\text{reio} \}$, we include 3 new parameters that capture the important effects of the new models: the DR contribution to \DNeff at late times, $\NIR$; the fraction of the dark matter energy density in the iDM component, $f_\chi = \rho_\chi/\rho_\dm$; and the redshift of the transition, $z_{t}$, defined as  the value of the redshift at which $T_d(z_t) = m_\psi$.\footnote{Throughout this paper we make the assumption that the DR is not present (or has a low enough temperature such that its energy density can be neglected) during BBN ($T_\gamma \sim 1~\MeV$), and that it is populated (or heated up) after this point. This assumption allows for larger values of $\DNeff$ to fix the $H_0$ problem, without at the same time increasing the primordial Helium fraction $Y_{\rm He}$ beyond what is permitted by BBN observations. This scenario can be easily realized in models where the DR has interactions with other energy density components, whose energy density gets transferred to the DR after the BBN era. In our code implementation of the \spa model, we simply set $\NUV = 0$ in the routine of the {\tt thermodynamics.c} module that computes $Y_{\rm He}$.} Of course, within \lcdm, these new parameters are simply $\NIR = 0$ and $f_\chi = 0$, with $\omega_\dm = \omega_\cdm$. For both \spa and \spap we use the flat priors $\NIR \geq 0$, $0 \leq f_\chi \leq 1$, and $4.0 \leq \log_{10} z_t \leq 5.5$\footnote{This relatively narrow prior on $\log z_t$ allows our scans to ignore the parameter space where the mass threshold takes place either very early (in which case the model becomes indistinguishable from that of self-interacting DR \cite{Blinov:2020hmc}) or very late (where the model becomes identical to PAcDM, \cite{Chacko:2015noa,Chacko:2016kgg,Buen-Abad:2017gxg}).}. We also use flat priors on the remaining \lcdm parameters. In all of our scans we include three active neutrinos, one with a mass of $0.06~\eV$ and the other two massless.

In principle, there are additional parameters, such as $m_\chi$ and $\alpha_d$; however, as discussed in~\cite{Buen-Abad:2022kgf}, these only enter the relevant equations through the combination $\alpha_d^2/m_\chi$, and thus we can keep $m_\chi = 1$ TeV fixed without any loss of generality. We will be interested in the regime in which the iDM--DR system begins its life as a single, strongly-interacting fluid. The precise redshift at which the $\psi$-$\chi$ scattering freezes-out only has a mild logarithmic dependence on $\alpha_d$, and therefore the precise value of the coupling $g_d$ is unimportant (we neglect small corrections to the tight coupling approximation). We adopt the benchmark value $\alpha_d = 10^{-3}$ in order to increase scanning speed and convergence. As with $g_d$, the precise value of the new coupling $g^{\prime}_d$ is also unimportant as long as it is within the range shown in \Eq{eq:alphad}.

For the generalized model \spap, there is a discrete choice for the extra number of fermion flavors we add to the dark fluid, $N_{\rm df}$. We focus on the scenarios with $N_{\rm df} \le 3$ since the impact of adding more flavors becomes smaller as $N_{\rm df}$ increases, see \Eq{eq:step-flavor}. While this means that we could, in principle, scan over an additional parameter $r_g$ which would correspond to \spa and \spap for special discrete values of $r_g$, this would lead to most of the points being scanned over representing non-physical realizations of the model. We therefore choose to treat each realization of the model separately. In the next section, we  present our numerical results only for $N_{\rm df} = 3$, which provides the best fit to the data.\footnote{We are unable to resist pointing out that 3 is also the number of generations in the visible sector.} 
We will make this choice explicit by referring to the model with $N_{\rm df}=3$ as \spapt.

\subsection{Experiments and Methodology}
\label{subsec:exps}

We perform a full likelihood analysis of the \spa and \spap models using various precision cosmological datasets. These include measurements of CMB anisotropies as well as galaxy, supernovae, and weak lensing surveys. We arrange these datasets into three distinct categories, identified with the abbreviations $\mD$, $\mH$ and $\mS$, based on their mutual compatibility within the context of \lcdm:

\begin{itemize}
    \item $\mD$: our baseline dataset. This includes the following experiments:
    \begin{itemize}
        \item {\bf Planck: } measurements of TT, TE, and EE CMB anisotropies and lensing from Planck 2018~\cite{Planck:2018vyg} (the likelihoods dubbed {\tt `Planck\_highl\_TTTEEE'}, {\tt `Planck\_lowl\_EE'}, {\tt `Planck\_lowl\_TT'}, and {\tt `Planck\_lensing'} in {\tt MontePython}).
        \item {\bf BAO: } baryon acoustic oscillations (BAO) datasets in the form of measurements of $D_V/r_\mathrm{drag}$ by the Six-degree Field Galaxy Survey (6dFGS) at $z = 0.106$ \cite{Beutler:2011hx} and by the Sloan Digital Sky Survey (SDSS) from the MGS galaxy sample at $z = 0.15$ \cite{Ross:2014qpa} ({\tt `bao\_smallz\_2014'}), as well as measurements of $D_M(z)/r_{s, \mathrm{drag}}$ and $H(z) r_{s, \mathrm{drag}}$ at $z = 0.38, \, 0.51, \, 0.61$ from the Baryon Oscillation Spectroscopic Survey (BOSS) DR12 \cite{BOSS:2016wmc} ({\tt `bao\_boss\_dr12'}).
        \item {\bf Pantheon: } measurements of the apparent magnitude of 1048 Type Ia supernovae (SNIa) at redshifts $0.01 < z < 2.3$, sampled by the Pan-STARRS1 collaboration \cite{Pan-STARRS1:2017jku} ({\tt `Pantheon'}).
    \end{itemize}
    \item $\mH$: the dataset containing late-universe measurements of the Hubble parameter $H_0$. While there are several of these measurements \cite{Freedman:2019jwv,Freedman:2021ahq,Yuan:2019npk,Soltis:2020gpl,Khetan:2020hmh,Huang:2019yhh,Schombert:2020pxm,Wong:2019kwg,Birrer:2020tax,Pesce:2020xfe,LIGOScientific:2019zcs,Abdalla:2022yfr}, at different levels of tension with results from \lcdm fits to $\mD$, we include the most precise:
    \begin{itemize}
        \item {\bf SH0ES: } the latest measurements of $H_0$ using Hubble Space Telescope (HST) observations of Cepheid variables in galaxies hosting 42 SNe Ia \cite{Riess:2021jrx}. The $H_0$ value quoted by the SH0ES collaboration is ultimately obtained from their derivation of the absolute magnitude $M_B$ of SNe Ia. We use their result, $M_B = -19.253 \pm 0.027$, to build a Gaussian likelihood.
    \end{itemize}
    \item $\mS$: the dataset with late-universe weak lensing and galaxy clustering measurements of $S_8 \equiv \sigma_8 \sqrt{\Omega_m / 0.3}$, which are in mild tension with the results from \lcdm fits to $\mD$. We include in this group the following experiments, with which we construct two-sided Gaussian likelihoods:
    \begin{itemize}
        \item {\bf DES: } the Year 3 results of the Dark Energy Survey (DES): $S_8 = 0.775^{+0.026}_{-0.024}$ \cite{DES:2021wwk}.
        \item {\bf KiDS-1000: } results from the Kilo-Degree Survey (KiDS-1000): $S_8 = 0.766^{+0.020}_{-0.014}$ \cite{Heymans:2020gsg}. 
    \end{itemize}
\end{itemize}
We denote any combination of these datasets by the corresponding abbreviation, such as $\mD \mH$ or $\mDHS$, following the notation of Ref.~\cite{Joseph:2022jsf}.

One could in principle include Lyman-$\alpha$ forest experiments, which are sensitive to even smaller scales and could therefore be very useful in distinguishing between various models that suppress the matter power spectrum. This data consists of the Lyman-$\alpha$ absorption lines present in the spectra of distant quasars (quasi-stellar objects, or QSOs), and arise from intergalactic neutral hydrogen lying along the quasars' line of sight. Indeed, since hydrogen traces the matter distribution (for redshifts of $2 \lesssim z \lesssim 5$ and length scales of $\mathcal{O}(1)~\Mpc/h \lesssim \lambda \lesssim \mathcal{O}(10)~\Mpc/h$ \cite{McQuinn:2015icp}), Lyman-$\alpha$ data can be used to probe the matter power spectrum. Some datasets commonly used in the literature include QSO spectra samples from SDSS-II \cite{McDonald:2004eu}, the HIRES/MIKE and XQ-100 experiments \cite{Viel:2013fqw,Irsic:2017ixq}, and from the BOSS and the Extended BOSS (eBOSS) collaborations~\cite{Chabanier:2018rga,BOSS:2012dmf,Dawson:2015wdb}. However, at present the usefulness of these experiments to constrain models beyond \lcdm that have an impact on the matter power spectrum is somewhat limited. In particular, some of these measurements rely on modeling the matter power spectrum in the context of \lcdm~\cite{McDonald:2004eu,McDonald:2004xn}, which may not be valid when considering models beyond \lcdm. Others require the modeling of various astrophysical nuisance parameters in conjunction with hydrodynamical simulations of structure formation \cite{Rossi:2014wsa,Borde:2014xsa,Palanque-Delabrouille:2015pga,Palanque-Delabrouille:2019iyz,Murgia:2018now}, which requires a prohibitively large amount of computer resources. Significant efforts to make the Lyman-$\alpha$ data friendlier to analysis of non-\lcdm models have recently been undertaken. In particular, the authors of Refs.~\cite{Murgia:2017lwo,Murgia:2017cvj,Murgia:2018now} introduced a description of the linear power spectrum in terms of distinct phenomenological ``shape parameters'' $\alpha$, $\beta$, and $\gamma$ (labelled the {\it ``$\{\alpha, \beta, \gamma \}$-parametrization''} in the literature), and used this parametrization, along with a suite of dedicated N-body simulations, to perform a MCMC analysis of the HIRES/MIKE and XQ-100 experiments and so derive a likelihood based on these datasets. This likelihood can then be easily used within {\tt MontePython} to analyze any non-\lcdm model whose linear matter power spectrum can be mapped into the three shape parameters employed in the analysis \cite{Archidiacono:2019wdp}. Despite its great versatility, this parametrization cannot be mapped to the power spectra from our \spa and \spap models, or those from the WZDR+ model of Ref.~\cite{Joseph:2022jsf}, which means that we cannot use this likelihood. Indeed, the $\{ \alpha, \beta, \gamma \}$-parametrization can only be used when the ratio of the linear matter power spectrum of a given model to that of \lcdm is suppressed down to zero for sufficiently large wavenumber $k$ (small length scales) in a power-law fashion. Within \spa (and \spap), the fact that $f_\chi < 1$ means that this ratio does not vanish completely, whereas in WZDR+ the suppression scales like $\sim \ln(k)$. While efforts to make the Lyman-$\alpha$ datasets applicable to a wider range of models are ongoing (see for example Ref.~\cite{Pedersen:2022anu}), no likelihood that can be easily applied to our model has been made publicly available at the time of the writing of this paper. Therefore, throughout the rest of this work, we limit ourselves to the cosmological observables described in the previous paragraphs, with the hope that in the future Lyman-$\alpha$ forest data can be used to improve our analysis.

\subsection{Numerical Results}
\label{subsec:num}

\begin{figure}[tb]
	\centering
	\includegraphics[width=.49\linewidth]{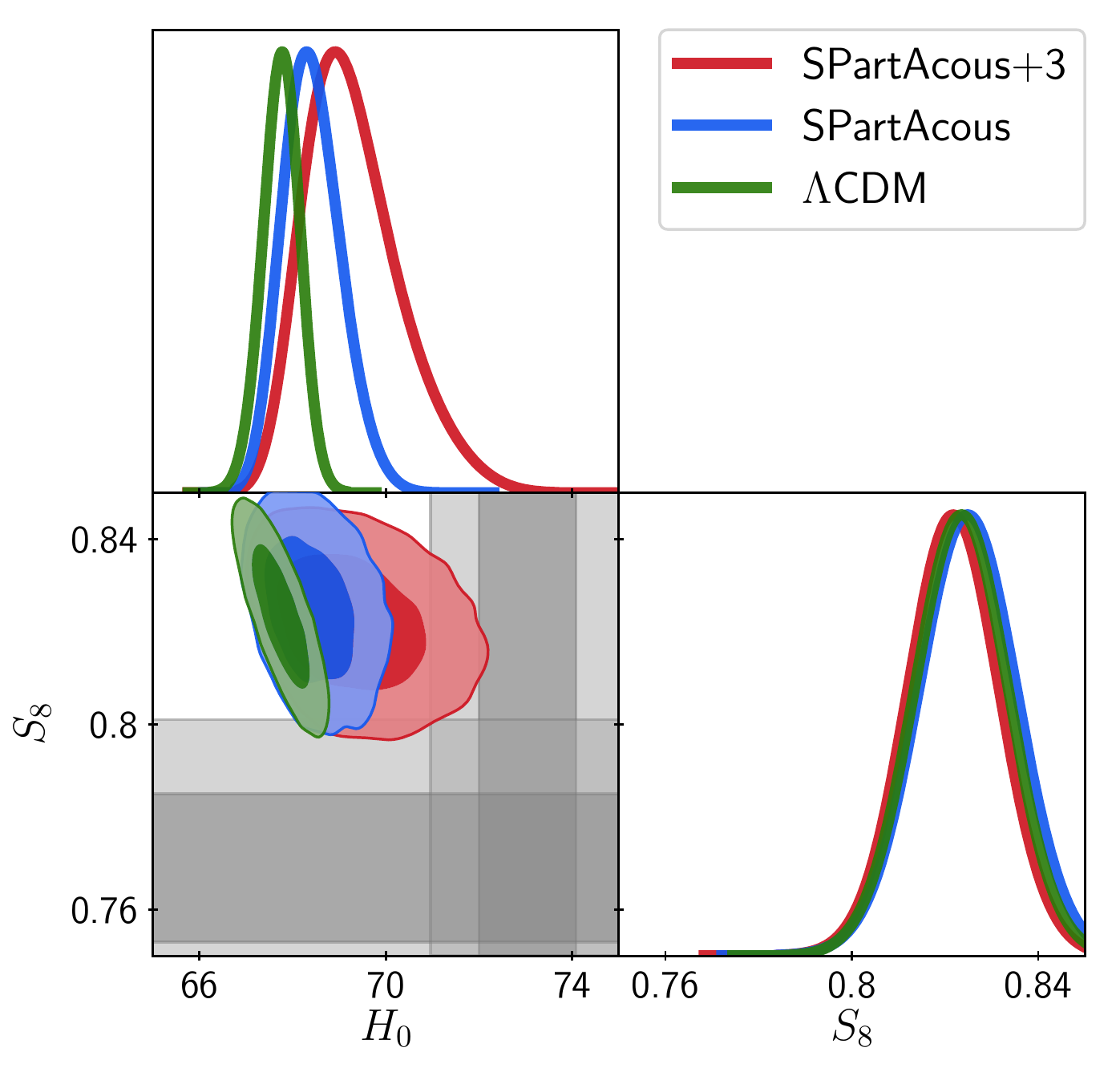}
	\includegraphics[width=.49\linewidth]{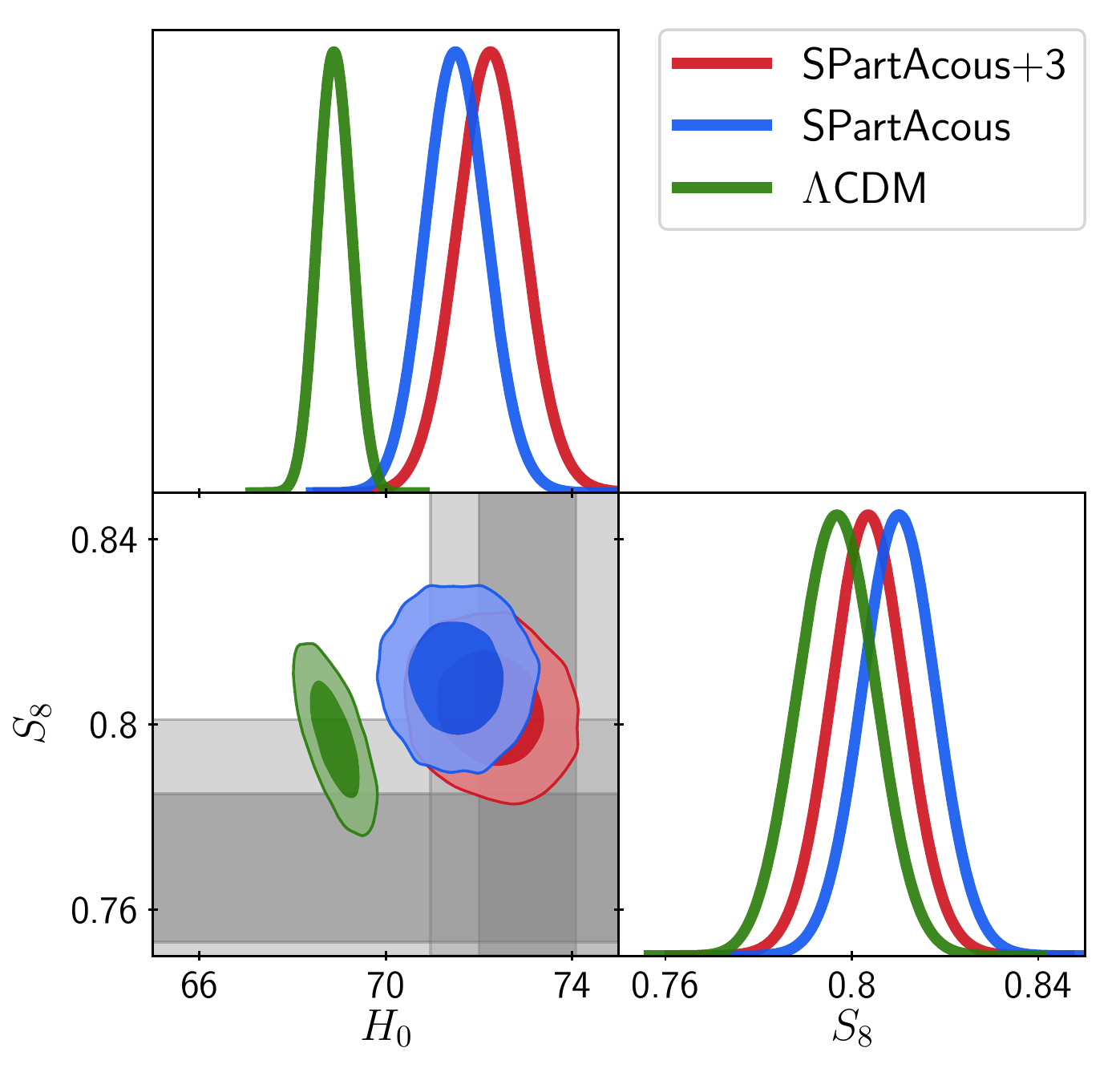}
	\caption{Posteriors for $H_0$ and $S_8$ for \lcdm ({\bf green} contours), \spa ({\bf blue} contours) and \spapt ({\bf red} contours) fit to the $\mD$ ({\bf left}) and $\mDHS$ ({\bf right}) datasets. The grey bands correspond to the 1 and 2 sigma regions for the local measurements of $H_0$ \cite{Riess:2021jrx} and $S_8$ \cite{Heymans:2020gsg}.}
	\label{fig:allmodelsDandDHS}
\end{figure}

\begin{figure}[tb]
	\centering
	\includegraphics[width=.49\linewidth]{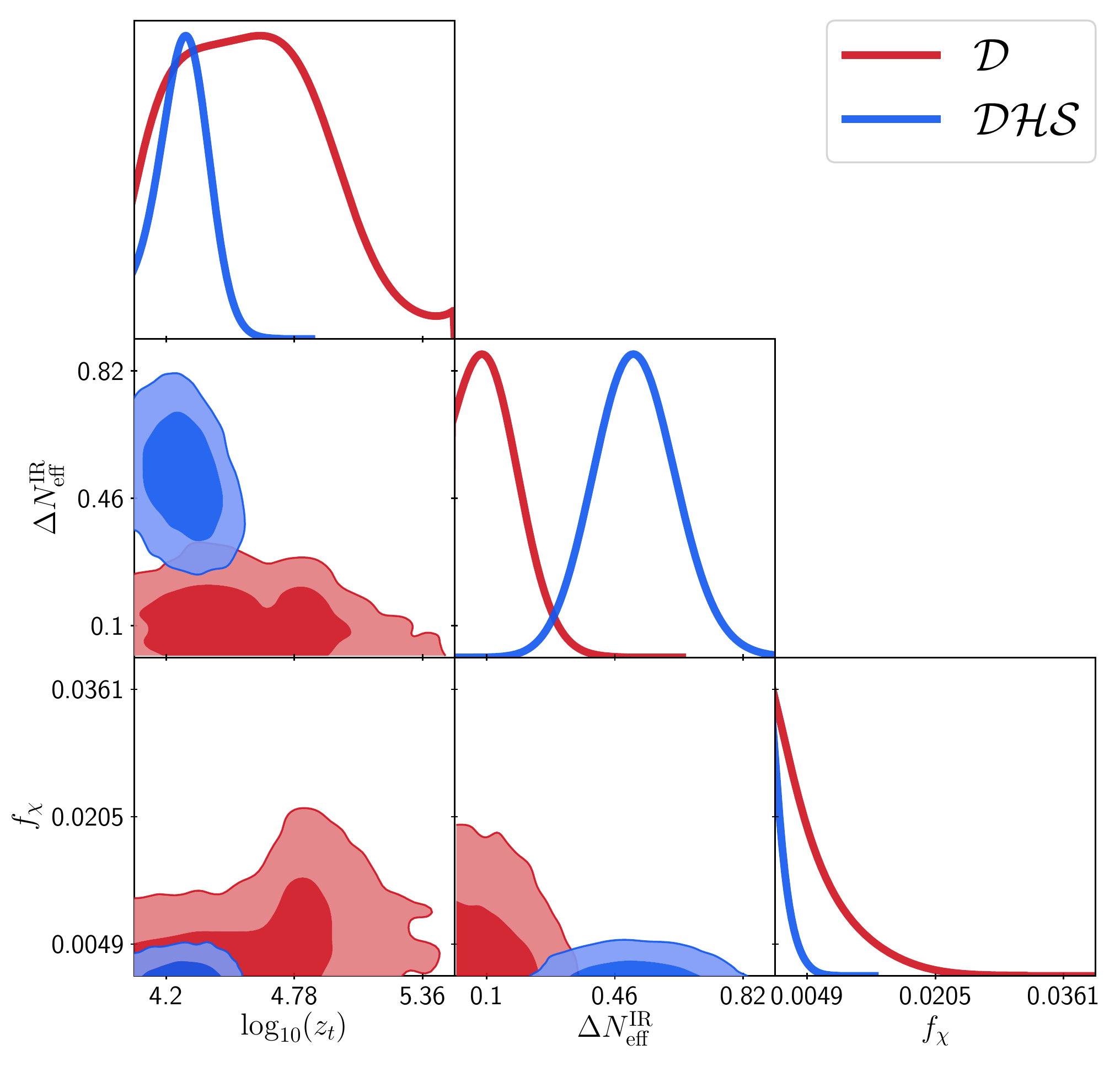}
	\includegraphics[width=.49\linewidth]{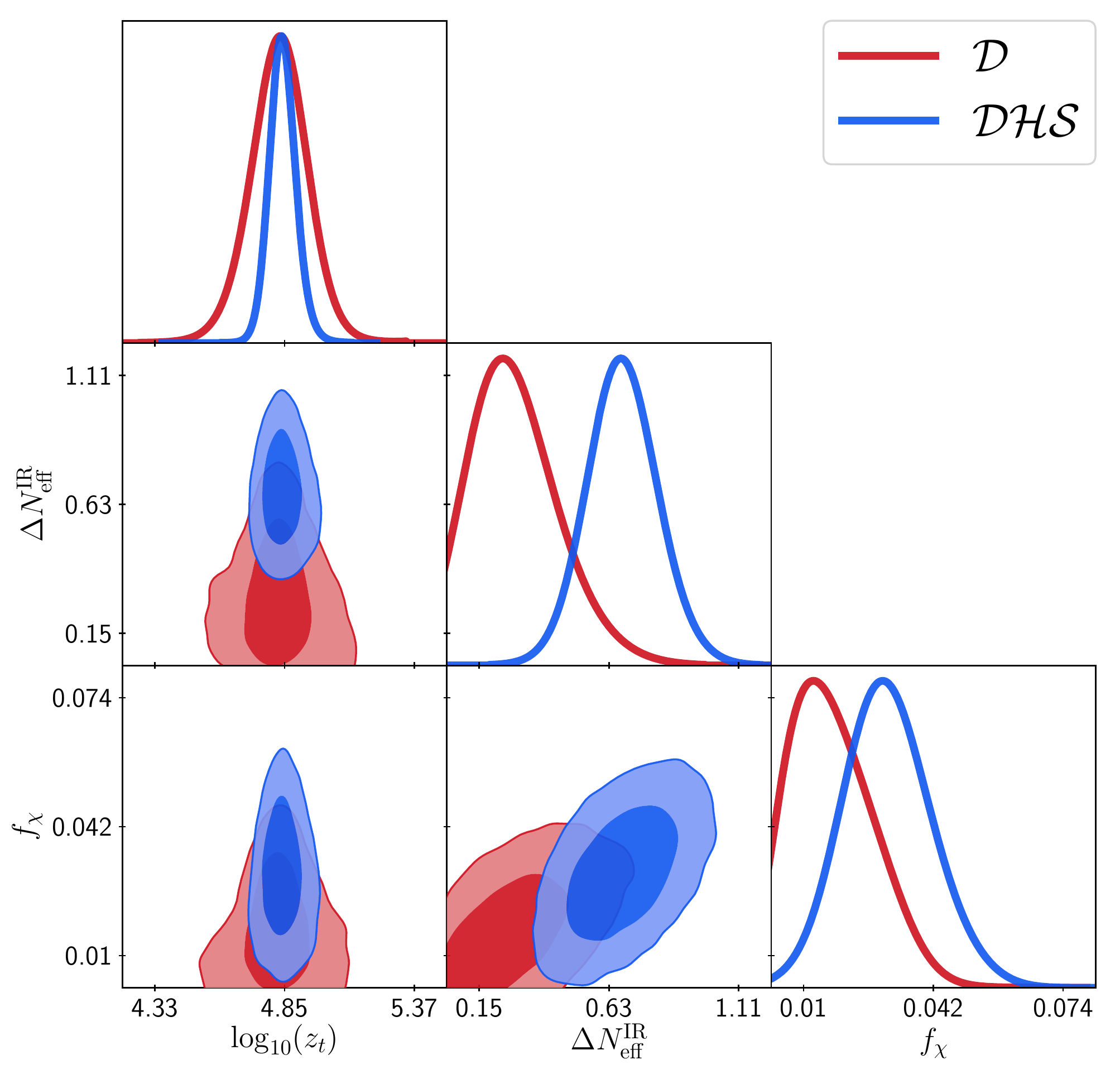}
	\caption{Posterior distributions for the new parameters in \spa ({\bf left}) and \spapt ({\bf right}) fit to the $\mD$ and $\mDHS$ datasets. The inclusion of the SH0ES likelihood pushes $N_{\rm IR}$ to larger values for both models, while $f_\chi$ becomes significantly constrained in \spa once it tries to accommodate a larger $H_0$.}
	\label{fig:2d-modelparameters}
\end{figure}

In Figure~\ref{fig:allmodelsDandDHS}, we show the marginalized posterior distributions for $H_0$ and $S_8$ for the three models under consideration, \lcdm, \spa and \spapt. The left panel shows the results when fitting to the $\mD$ dataset, while the right panel shows the results when fitting to the $\mDHS$ data combination. We can immediately see that, even for the fit to $\mD$, both \spa and \spapt allow for larger values of $H_0$ compared to \lcdm. In particular, for \spapt, we see that, even though it does not completely solve the tension, the 95\% credible region (CR) overlaps with the $1\sigma$ region from the most recent SH0ES measurement of $H_0$~\cite{Riess:2021jrx}. The allowed ranges for $S_8$ are approximately the same for all three models when fitting to the $\mD$ dataset. The posteriors for the fit to $\mDHS$, on the right side of the figure, show bigger differences between the 3 models. For \lcdm, we see that the inclusion of the extra data lowers the value of $S_8$, while it leads to only mild changes to $H_0$, despite the fact that $H_0$ is a more significant tension. This illustrates that there is significant tension between the different data sets in \lcdm. For \spa, we see that with the inclusion of the SH0ES prior, it can accommodate much larger $H_0$ values, but it is not able to simultaneously lower $S_8$. In fact, we find that when the extra data is included, the allowed range for $f_\chi$, which controls the decrease in the power spectrum at small scales, becomes very small and peaked at zero iDM, as can be seen in Figure~\ref{fig:2d-modelparameters}. As we will see later, this is driven by the inclusion of $\mathcal{H}$. On the other hand, we see that with \spapt we can achieve even larger values of $H_0$, while simultaneously lowering $S_8$. This was the expectation for this class of interacting DR models, in which increasing $\DNeff$ allows higher $H_0$ while increasing $f_\chi$ reduces $S_8$.

\begin{table}[th!]
\centering
\begin{adjustbox}{max width=\columnwidth}
\begin{tabular}{|c|cc|cc|cc|}
\toprule\toprule
Model                         & \multicolumn{2}{c|}{$\lcdm$}                         & \multicolumn{2}{c|}{\spa}                      & \multicolumn{2}{c|} \spapt                    \\ \midrule
Dataset                       & \multicolumn{1}{c|}{$\mD$} & $\mDHS$ & \multicolumn{1}{c|}{$\mD$} & $\mDHS$ & \multicolumn{1}{c|}{$\mD$} & $\mDHS$ \\ \midrule
$100~\theta_s$                 & \multicolumn{1}{c|}{$1.042$}       & $1.042$         & \multicolumn{1}{c|}{$1.042$}       & $1.043$         & \multicolumn{1}{c|}{$1.043$}       & $1.044$         \\
$100~\omega_{b}$              & \multicolumn{1}{c|}{$2.246$}       & $2.270$         & \multicolumn{1}{c|}{$2.251$}       & $2.277$         & \multicolumn{1}{c|}{$2.268$}       & $2.317$         \\
$\omega_\dm$          & \multicolumn{1}{c|}{$0.1192$}      & $0.1168$        & \multicolumn{1}{c|}{$0.1213$}      & $0.1255$        & \multicolumn{1}{c|}{$0.1250$}      & $0.1327$        \\
$\ln 10^{10} A_s$             & \multicolumn{1}{c|}{$3.047$}       & $3.050$         & \multicolumn{1}{c|}{$3.048$}       & $3.046$         & \multicolumn{1}{c|}{$3.050$}       & $3.045$         \\
$n_s$                         & \multicolumn{1}{c|}{$0.9682$}      & $0.9743$        & \multicolumn{1}{c|}{$0.9720$}       & $0.9848$        & \multicolumn{1}{c|}{$0.9737$}      & $0.9822$        \\
$\tau_\mathrm{reio}$          & \multicolumn{1}{c|}{$0.0558$}      & $0.05967$       & \multicolumn{1}{c|}{$0.05583$}     & $0.05585$       & \multicolumn{1}{c|}{$0.05655$}     & $0.05674$       \\
$\NIR$               & \multicolumn{1}{c|}{$-$}           & $-$             & \multicolumn{1}{c|}{$0.12$}        & $0.51$          & \multicolumn{1}{c|}{$0.25$}        & $0.69$          \\
$f_\chi [\%]$                      & \multicolumn{1}{c|}{$-$}           & $-$             & \multicolumn{1}{c|}{$0.0$}       & $0.0$         & \multicolumn{1}{c|}{$1.7$}       & $3.3$         \\
$\log_{10}(z_t)$              & \multicolumn{1}{c|}{$-$}           & $-$             & \multicolumn{1}{c|}{$4.36$}        & $4.26$          & \multicolumn{1}{c|}{$4.81$}        & $4.84$          \\ \midrule
$M_B$                         & \multicolumn{1}{c|}{$-19.415$}     & $-19.384$       & \multicolumn{1}{c|}{$-19.392$}     & $-19.305$       & \multicolumn{1}{c|}{$-19.371$}     & $-19.279$       \\
$H_0~[\km/\seg/\Mpc]$              & \multicolumn{1}{c|}{$67.79$}       & $68.94$         & \multicolumn{1}{c|}{$68.46$}       & $71.55$         & \multicolumn{1}{c|}{$69.07$}       & $72.26$         \\
$\sigma_8$                    & \multicolumn{1}{c|}{$0.8099$}      & $0.8041$        & \multicolumn{1}{c|}{$0.8156$}       & $0.8228$        & \multicolumn{1}{c|}{$0.808$}       & $0.8039$        \\
$S_8$                         & \multicolumn{1}{c|}{$0.8227$}      & $0.7972$        & \multicolumn{1}{c|}{$0.8266$}      & $0.8103$        & \multicolumn{1}{c|}{$0.8224$}      & $0.8036$        \\ \midrule\midrule
$\chi^2_\mathrm{CMB}$         & \multicolumn{1}{c|}{$2765.80$}     & $2772.08$       & \multicolumn{1}{c|}{$2765.06$}     & $2768.68$       & \multicolumn{1}{c|}{$2764.02$}     & $2769.39$       \\
$\chi^2_\mathrm{Pantheon}$    & \multicolumn{1}{c|}{$1025.86$}     & $1025.78$       & \multicolumn{1}{c|}{$1025.94$}     & $1025.85$       & \multicolumn{1}{c|}{$1026.38$}     & $1025.78$       \\
$\chi^2_\mathrm{BAO}$         & \multicolumn{1}{c|}{$5.48$}        & $6.37$          & \multicolumn{1}{c|}{$5.30$}        & $7.88$          & \multicolumn{1}{c|}{$5.61$}        & $5.47$          \\
$\chi^2_\mathrm{Pl. lensing}$ & \multicolumn{1}{c|}{$8.84$}        & $10.10$         & \multicolumn{1}{c|}{$9.09$}        & $10.86$         & \multicolumn{1}{c|}{$8.89$}        & $10.03$         \\
$\chi^2_{S_8}$                & \multicolumn{1}{c|}{$-$}           & $3.16$          & \multicolumn{1}{c|}{$-$}           & $6.75$          & \multicolumn{1}{c|}{$-$}           & $4.74$          \\
$\chi^2_\mathrm{SH0ES}$       & \multicolumn{1}{c|}{$-$}           & $23.40$         & \multicolumn{1}{c|}{$-$}           & $3.68$          & \multicolumn{1}{c|}{$-$}           & $0.93$          \\ \midrule
$\chi^2_\mathrm{tot}$         & \multicolumn{1}{c|}{$3805.98$}     & $3840.90$       & \multicolumn{1}{c|}{$3805.39$}     & $3823.70$       & \multicolumn{1}{c|}{$3804.90$}     & $3816.34$       \\ \bottomrule\bottomrule
\end{tabular}
\end{adjustbox}
\caption{Best-fit values of the parameters of the \lcdm, \spa, and \spapt ($N_{\rm df} = 3$) models, fitted to datasets $\mD$ and $\mDHS$.}
\label{tab:bf_ddhs}
\end{table}

In Table~\ref{tab:bf_ddhs}, we show the best-fit points for the parameters of the three models when fit to $\mD$ and $\mDHS$, as well as their corresponding $\chi^2$, broken down by the contributions from each dataset. Note that when fit to $\mD$, both \spa and \spapt improve the $\chi^2$, although only by a limited amount considering that these models have 3 extra parameters compared to \lcdm. However, once we include $\mathcal{H}$ and $\mathcal{S}$, we see that both models lead to a significant improvement over \lcdm, with \spapt giving $\Delta \chi^2 = -24.6$. Even when taking into account the penalty for having three extra parameters ($\NIR$, $f_\chi$, and $\log_{10}(z_t)$) by using the Akaike Information Criterion (AIC), the improvement remains substantial. Indeed, from ${\rm AIC} \equiv \chi^2_{\rm b.f.} + 2 \times {\rm d.o.f.}$, we find $\Delta {\rm AIC} = -18.6 \, (-11.2)$ for fits of \spapt (\spa) to $\mDHS$; see Table~\ref{tab:dhs}. By contrast, fits to only $\mD$ are disfavored by the AIC as can be seen in Table~\ref{tab:d}. In fact, note that once the $\mathcal{H}$ and $\mathcal{S}$ data are included in the fit, \spa and \spapt provide an improved $\chi^2$ compared to \lcdm even when computing only the contribution coming from CMB data. The results in Table~\ref{tab:bf_ddhs} also show that \spapt provides an overall better fit than \spa, and as expected from Figure~\ref{fig:allmodelsDandDHS}, we see that \spapt provides a more significant reduction of both the $H_0$ and $S_8$ tensions. As expected, the inclusion of the $\mH$ prior pushes $H_0$ to higher values, in both \lcdm and the \spa and \spapt extensions. Some important points to note are that: ({\it i.}) \spa and \spapt allow for larger values for $H_0$ than \lcdm with {\it or} without the $\mH$ prior, ({\it ii.}) the overall global fit is better in \spa and \spapt than in \lcdm, even accounting for the extra degrees of freedom, and ({\it iii.}) the improvement in $\chi_{\rm SH0ES}^2$ in the \spa and \spap models does {\it not} come at the cost of a degradation of the $\chi^2_\mathrm{CMB}$, which is lower in both models compared to \lcdm when fitting to $\mDHS$.

We present the $95\%$ CR range for the parameters $\NIR$, $f_\chi$, $H_0$, and $S_8$ in Table~\ref{tab:d} for the fit to $\mD$ and in Table~\ref{tab:dhs} for the fit to $\mDHS$. The posteriors for the new parameters, $\NIR$, $f_\chi$, and $\log z_t$, are also shown in Figure~\ref{fig:2d-modelparameters} (see Appendix~\ref{sec:appnum} for the triangle plots with the posteriors of all parameters). One can see from Table~\ref{tab:d} that both classes of models allow for much larger values of $\DNeff$, even without including the SH0ES results, compared to simple extensions of \lcdm where DR is purely free-streaming~\cite{Planck:2018vyg} or purely interacting~\cite{Blinov:2020hmc}. These larger values for \DNeff lead to larger values for $H_0$; in particular, \spapt yields a value of $H_0$ as large as $71.4 \, \mathrm{km/s/Mpc}$ at 95\% CR, compared to at most $68.6 \, \mathrm{km/s/Mpc}$ for \lcdm. This table also shows that, in \spa, the maximum allowed $f_\chi$ is much smaller than in \spapt, which explains why \spapt is more effective in addressing the $S_8$ tension. In Tables~\ref{tab:ms_d} and~\ref{tab:ms_dhs}, we show the mean and $1\sigma$ allowed ranges for all the parameters in the three models, fit to $\mD$ and $\mDHS$ respectively. We can see that when fitting to $\mD$, the main difference in the parameters common to all models is that in both \spa and \spap, the allowed ranges for $\omega_\text{dm}$ and $n_s$ are widened and the central value pushed to slightly larger values to compensate for the inclusion of extra dark radiation, as has been the case in other models with additional radiation~\cite{Blinov:2020hmc,Aloni:2021eaq}. This is expected in order to keep the redshift at matter radiation equality fixed, and also to compensate for the change in Silk damping. In the fit to $\mDHS$, due to the pressure from the SH0ES data to increase $H_0$, the mean for $\DNeff$ is significantly increased. This also pushes the mean values of $\omega_\text{dm}$ and $n_s$ to larger values.

\begin{table}[h!]
\centering
\begin{adjustbox}{max width=\columnwidth}
\begin{tabular}{|c|c|c|c|c|c|c|}
\toprule
Model        &  $\Delta\chi^2$ & $\Delta {\rm AIC}$ & $\NIR$ & $f_\chi [\%]$ & $H_0~[\km/\seg/\Mpc]$ & $S_8$\\ \midrule
$\lcdm$      &  -- & -- & $3.04$              & -- & $ [67.0,68.6]$ &  [0.80,0.84] \\ 
\spa   & $-0.59$ & $+5.41$ & $[3.04,3.31]$  &  [0.0, 1.5] & $ [67.3,69.7]$ &  [0.80,0.85] \\ 
\spapt & $-1.08$ & $+4.92$ & $[3.04,3.63]$  & [0.0, 3.6] & $ [67.4,71.4]$ &  [0.80,0.84]\\ \bottomrule
\end{tabular}
\end{adjustbox}
\caption{A summary of the fits to the dataset $\mD$ for the \lcdm, \spa, and \spapt models, showing the allowed parameter range at 95\% CR for $\NIR$, $f_\chi$ and $H_0$. These intervals are defined to be the narrowest interval containing 95\% of the integrated posterior density, and have been computed directly from the posterior densities and not using Gaussian fits to the posteriors.}
\label{tab:d}
\end{table}

\begin{table}[h!]
\centering
\begin{adjustbox}{max width=\columnwidth}
\begin{tabular}{|c|c|c|c|c|c|c|}
\toprule
Model        &  $\Delta\chi^2$ & $\Delta {\rm AIC}$ & $\NIR$ & $f_\chi [\%]$ & $H_0~[\km/\seg/\Mpc]$   & $S_8$              \\ \midrule
$\lcdm$      & -- & -- & $3.04$              & --            & $[68.2,69.6]$ & $[0.78,0.81]$ \\ 
\spa   & $-17.20$ & $-11.20$ & $[3.33,3.79]$       &  [0.0, 0.4]   & $[70.2,72.9]$ & $[0.79,0.83]$ \\ 
\spapt & $-24.56$ & $-18.56$ & $[3.46,3.99]$       &  [0.9, 5.2]   & $[70.8,73.7]$ & $[0.79,0.82]$ \\ \bottomrule
\end{tabular}
\end{adjustbox}
\caption{A summary of the fits to the dataset $\mDHS$ for the \lcdm, \spa, and \spapt models, showing the allowed parameter range at 95\% CR for $\NIR$, $f_\chi$ and $H_0$. These intervals are defined to be the narrowest interval containing 95\% of the integrated posterior density, and have been computed directly from the posterior densities and not using Gaussian fits to the posteriors.}
\label{tab:dhs}
\end{table}

\begin{table}[h!]
\centering
\begin{adjustbox}{max width=\columnwidth}
\begin{tabular}{|c|c|c|c|}
\toprule
Model                & $\lcdm$                                           & \spa                                        & \spapt                                      \\ \midrule
$100~\theta_s$        & $1.0420_{-0.0003}^{+0.0003}$                     & $1.0423_{-0.0004}^{+0.0003}$                     & $1.0427_{-0.0005}^{+0.0004}$                     \\
$100~\omega_{b}$     & $2.243_{-0.014}^{+0.013}$                         & $2.252_{-0.015}^{+0.015}$                         & $2.271_{-0.021}^{+0.018}$                         \\
$\omega_\dm$ & $0.1192_{-0.0009}^{+0.0009}$                     & $0.1216_{-0.0020}^{+0.0014}$                       & $0.1257_{-0.0043}^{+0.0028}$                      \\
$\ln 10^{10} A_s$    & $3.047_{-0.014}^{+0.013}$                         & $3.050_{-0.015}^{+0.013}$                          & $3.051_{-0.015}^{+0.013}$                         \\
$n_s$                & $0.9666_{-0.0033}^{+0.0034}$                      & $0.9722_{-0.0046}^{+0.0042}$                      & $0.9735_{-0.0051}^{+0.0044}$                      \\
$\tau_\mathrm{reio}$ & $0.05627_{-0.00721}^{+0.00654}$                     & $0.05668_{-0.00720}^{+0.00631}$                     & $0.05713_{-0.00688}^{+0.00644}$                     \\
$\NIR$      & $-$                                               & $0.1205_{-0.1194}^{+0.0331}$                         & $0.2892_{-0.2}^{+0.12}$                           \\
$f_\chi [\%]$             & $-$                                               & $0.5216_{-0.5216}^{+0.1090}$                        & $1.7477_{-1.2927}^{+0.7382}$                          \\
$\log_{10}(z_t)$     & $-$                                               & $4.581_{-\mathrm{nan}}^{+\mathrm{nan}}$                              & $4.825_{-0.094}^{+0.085}$                         \\ \midrule
$M_B$                & $-19.416_{-0.011}^{+0.011}$                        & $-19.394_{-0.021}^{+0.016}$                        & $-19.365_{-0.037}^{+0.026}$                        \\
$H_0~[\km/\seg/\Mpc]$     & $67.75_{-0.41}^{+0.41}$                           & $68.42_{-0.70}^{+0.56}$                            & $69.34_{-1.20}^{+0.84}$                            \\
$\sigma_8$           & $0.8099_{-0.0059}^{+0.0056}$ & $0.8122_{-0.0072}^{+0.0075}$ & $0.8086_{-0.0073}^{+0.0073}$ \\
$S_8$                & $0.8233_{-0.0102}^{+0.0105}$                          & $0.8247_{-0.0106}^{+0.0104}$                         & $0.8219_{-0.0100}^{+0.0102}$                          \\ \bottomrule
\end{tabular}
\end{adjustbox}
\caption{Mean and $\pm 1\sigma$ values of the fits of the \lcdm, \spa, and \spapt models to the dataset $\mD$. Note that the $\pm 1 \sigma$ values of the $\log_{10} (z_t)$ parameter in \spa are ``nan''. This is because of the very non-Gaussian nature of its posterior.}
\label{tab:ms_d}
\end{table}

\begin{table}[htb]
\centering
\begin{adjustbox}{max width=\columnwidth}
\begin{tabular}{|c|c|c|c|}
\toprule
Model                & $\lcdm$            & \spa                                        & \spapt                                      \\ \midrule
$100~\theta_s$        & $1.0422_{-0.0003}^{+0.0003}$  & $1.0431_{-0.0003}^{+0.0003}$                     & $1.0436_{-0.0005}^{+0.0005}$                     \\
$100~\omega_{b}$     & $2.267_{-0.013}^{+0.013}$      & $2.275_{-0.015}^{+0.014}$                         & $2.316_{-0.016}^{+0.016}$                         \\
$\omega_\dm$ & $0.1168_{-0.0008}^{+0.0008}$ & $0.1259_{-0.0022}^{+0.0020}$                       & $0.1323_{-0.0035}^{+0.0032}$                      \\
$\ln 10^{10} A_s$    & $3.049_{-0.015}^{+0.014}$      & $3.047_{-0.013}^{+0.012}$                         & $3.045_{-0.014}^{+0.013}$                         \\
$n_s$                & $0.9727_{-0.0032}^{+0.0032}$   & $0.9848_{-0.0043}^{+0.0044}$                      & $0.9817_{-0.0047}^{+0.0046}$                      \\
$\tau_\mathrm{reio}$ & $0.05957_{-0.00796}^{+0.00685}$   & $0.05594_{-0.00664}^{+0.00599}$                      & $0.05715_{-0.00709}^{+0.00626}$                     \\
$\NIR$      & $-$                            & $0.5182_{-0.1212}^{+0.1136}$                          & $0.6802_{-0.14}^{+0.13}$                          \\
$f_\chi [\%]$             & $-$                            & $0.1453_{-0.1453}^{+0.0317}$                       & $3.0854_{-1.1224}^{+1.0788}$                           \\
$\log_{10}(z_t)$     & $-$                            & $4.268_{-0.110}^{+0.120}$                           & $4.841_{-0.056}^{+0.046}$                         \\ \midrule
$M_B$                & $-19.383_{-0.010}^{+0.009}$   & $-19.305_{-0.019}^{+0.019}$                         & $-19.280_{-0.022}^{+0.022}$                        \\
$H_0$ [km/Mpc/s]     & $68.91_{-0.36}^{+0.35}$        & $71.54_{-0.67}^{+0.66}$                           & $72.25_{-0.76}^{+0.75}$                           \\
$\sigma_8$           & $0.8034_{-0.0059}^{+0.0053}$   & $0.8218_{-0.0065}^{+0.0065}$ & $0.8047_{-0.0071}^{+0.0069}$ \\
$S_8$                & $0.7968_{-0.0083}^{+0.0083}$   & $0.8103_{-0.0081}^{+0.0079}$                      & $0.8035_{-0.0083}^{+0.0080}$                       \\ \bottomrule
\end{tabular}
\end{adjustbox}
\caption{Mean and $\pm 1\sigma$ values of the fits of the \lcdm, \spa, and \spapt models to datasets $\mDHS$.}
\label{tab:ms_dhs}
\end{table}

\subsubsection{Fit to $\mathcal{DH}$ and Cosmic Concordance}

\begin{figure}[tb]
	\centering
	\includegraphics[width=.49\linewidth]{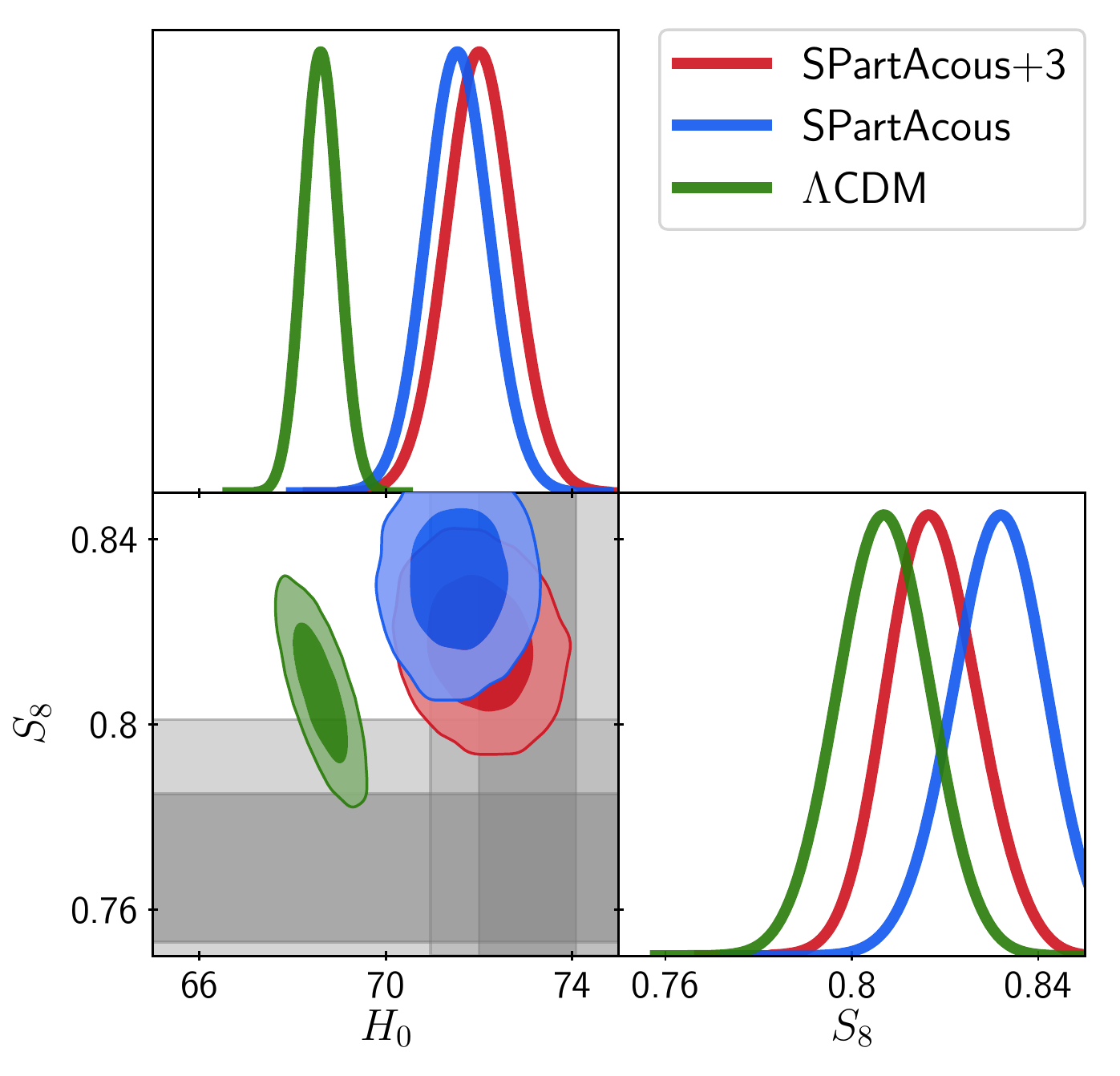}
	\caption{Posterior distributions for $H_0$ and $S_8$ for $\lcdm$ ({\bf green}), \spa ({\bf blue}) and \spapt ({\bf red}) fit to the $\mDH$ dataset. Compared to the fit to the $\mD$, note that both \spa and \spapt get pulled to larger values for $H_0$, while $\lcdm$ sees almost no change.}
	\label{fig:2d-DH}
\end{figure}

Currently, the $H_0$ tension is much more significant than $S_8$. For this reason, we have also made runs including only the $\mDH$ data combination with all three models. One can see in Figure~\ref{fig:2d-DH} that the inclusion of the SH0ES prior pulls $H_0$ to larger values in both \spa and \spapt, as expected, while it only leads to a very moderate shift for \lcdm. In \spa, the allowed range for $S_8$ gets pushed to larger values compared to the results fit to $\mD$ shown in Figure~\ref{fig:allmodelsDandDHS}, which is a reflection of the fact that the allowed range for iDM in \spa is very small and cannot efficiently lower $S_8$. On the other hand, in \spapt, we see that despite the significant increase in $H_0$ values, the allowed range for $S_8$ remains almost unchanged compared to the fit to $\mD$. This is in contrast to many of the most prmoising proposals to address the $H_0$ tension~(see e.g. \cite{Schoneberg:2021qvd}).

In order to quantify whether the models are compatible with the inclusion of both datasets, we will follow the approach used in~\cite{Schoneberg:2019wmt}~(see \cite{Raveri:2018wln} for more in depth discussion) and compute the $Q_\text{DMAP}$ of a given model fit to the two datasets, where
\beq
    Q_\text{DMAP}(\text{Model X}) = \sqrt{\chi^2_{\text{min},\mDH} - \chi^2_{\text{min},\mD}} \, .
\eeq
This quantity is generally interpreted as the number of standard deviations (i.e. $\sigma$) as a measure of the tension between the two datasets. For the datasets to be compatible within the context of a given model this quantity is expected to be small; a large value indicates that, for the model under study, there is tension between these datasets. In Table~\ref{tab:qdmap}, we show the $\chi^2_\text{min}$ for the fits to $\mDH$ and the corresponding $Q_\text{DMAP}$ values for the three models under consideration. We see that there is significant improvement in the tension in both \spa and \spapt compared to \lcdm, going from $\sim 5 \sigma$ tension in \lcdm to $\sim 2 \sigma$ in \spapt. The improvement achieved with \spapt is similar to the one recently found in Ref.~\cite{Joseph:2022jsf} using WZDR+, although their definition of $\mathcal{D}$ does not include the lensing likelihood, and so a direct comparison of the results is not possible. It is also similar to the results obtained by the models ranked highest in the $H_0$ Olympics~\cite{Schoneberg:2021qvd}, showing that \spa and \spapt are amongst the most successful proposals to solve the $H_0$ tension.

\begin{table}[th]
    \centering
    \begin{adjustbox}{max width=\columnwidth}
    \begin{tabular}{|c|c|c|}
    \toprule
      Model   & $\chi^2_{\text{min},\mDH}$ & $Q_\text{DMAP}$ \\ \midrule
      \lcdm   & $3836.57$                   & 5.53 \\
      \spa   &  $3813.33$           & 2.82 \\
      \spap   & $3809.68$           & 2.19 \\ \bottomrule
    \end{tabular}
    \end{adjustbox}
    \caption{List of the $\chi^2$ values for the best-fit points of the \lcdm, \spa, and \spapt models when fitting to $\mDH$, and their corresponding $Q_\text{DMAP}$ when comparing to the fit to $\mD$.}
    \label{tab:qdmap}
\end{table}

\subsubsection{Matter Power Spectrum}

Given the significant improvement in the fit with \spapt, an important question is whether future measurements can distinguish it from other models that address the \HO and \Se tensions. Given the large number of solutions that have been proposed, a comprehensive analysis is beyond the scope of this work. We will therefore limit the discussion to a comparison with two other early universe (pre-recombination) solutions of the $H_0$ tension, Early Dark Energy (EDE) \cite{Poulin:2016nat,Smith:2019ihp}, WZDR, and its generalizations \cite{Aloni:2021eaq,Joseph:2022jsf}. For EDE and for the original WZDR \cite{Aloni:2021eaq} proposal, increasing $H_0$ correlates with an increase in $S_8$, and therefore future measurements of $S_8$ with increasing precision should clearly distinguish between the two models. In fact, due to the present $S_8$ tension, these models are already in some tension with more direct measurements of the matter power spectrum. The recently proposed WZDR+ model, which includes very weak interactions of all of dark matter with the DR, was considered in Ref.~\cite{Joseph:2022jsf}~(see also Ref.~\cite{Schoneberg:2022grr} for a similar construction, but with an overall worse fit to $\mDHS$). As in \spapt, the interactions between dark matter and DR can suppress the power spectrum at small scales and address the $S_8$ tension. Nonetheless, even if both \spapt and WZDR+ models lead to similar results for $S_8$, that is but a single number; their overall impact on the full power spectrum is very different.

\begin{figure}[tb]
	\centering
	\includegraphics[width=.49\linewidth]{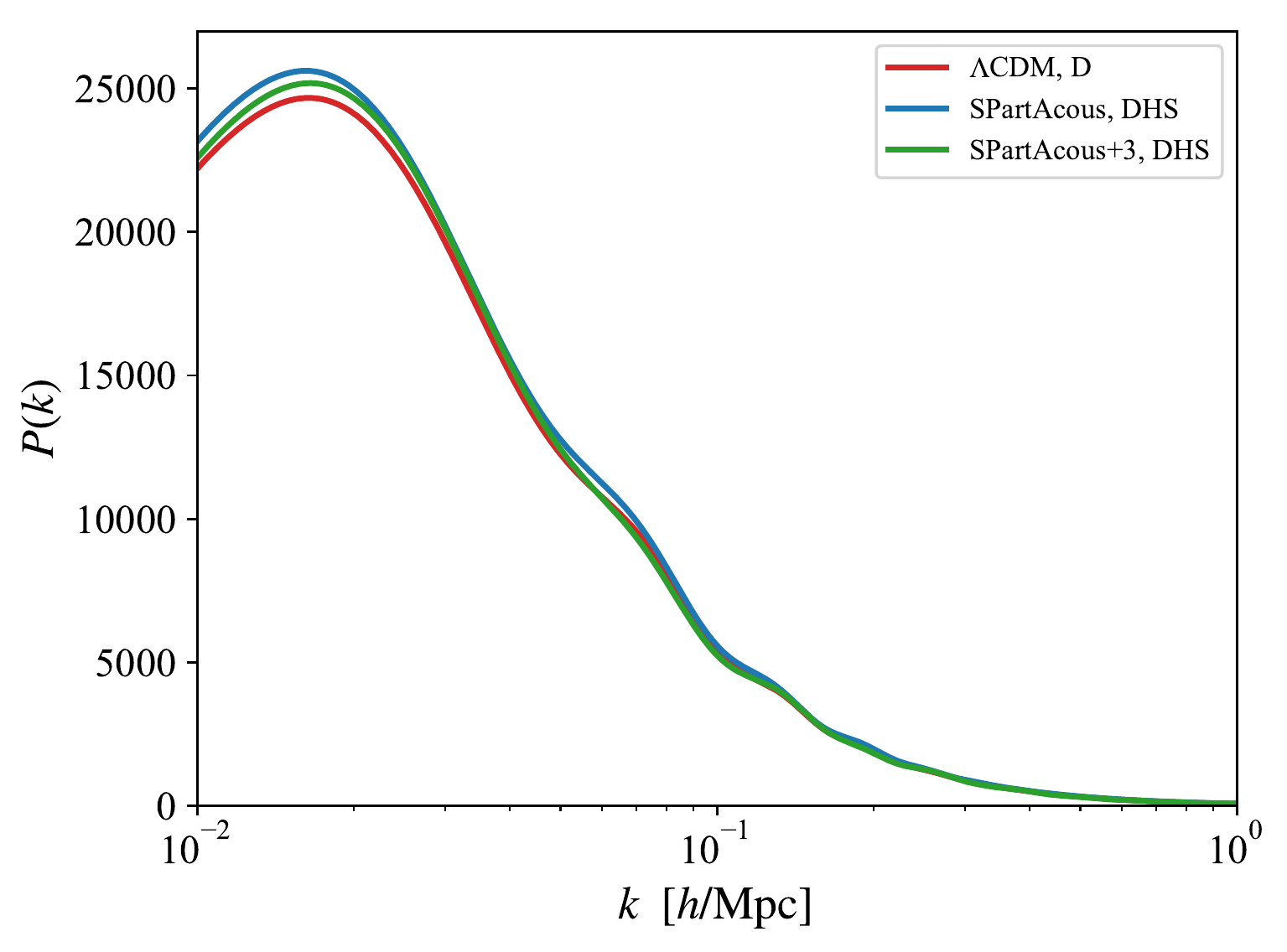}
    \includegraphics[width=.49\linewidth]{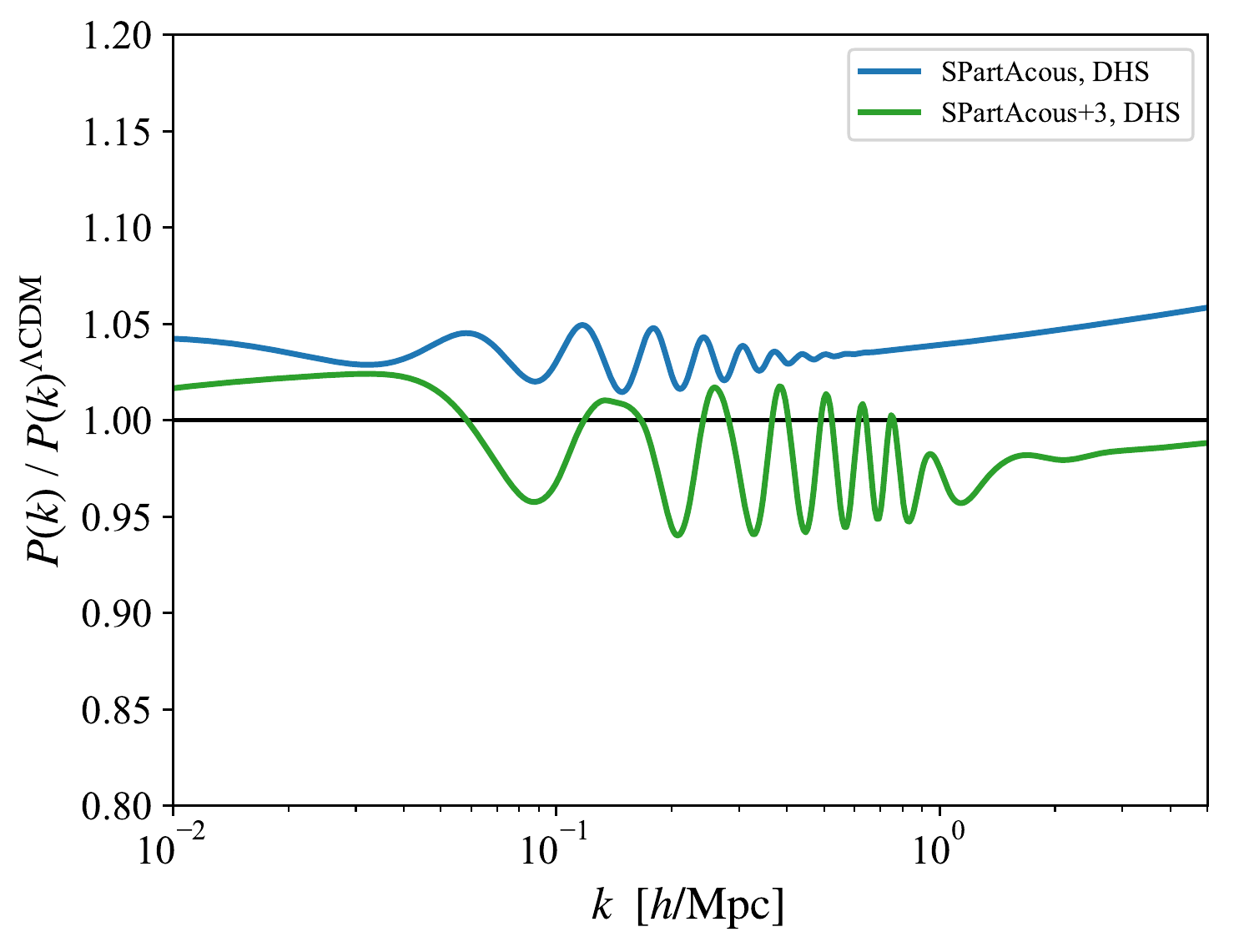}
	\caption{{\bf Left}: Matter power spectrum for \lcdm using the best fit point to $\mD$ and that for \spa and \spapt using the best fit point to the $\mDHS$ data.   {\bf Right}: Ratio of the matter power spectrum of \spa and \spapt using the best fit point to the $\mDHS$ to that of \lcdm using the best fit point to $\mD$.}
	\label{fig:Pk}
\end{figure}

In Figure~\ref{fig:Pk}, we show the matter power spectrum for our best-fit models. As discussed in Ref.~\cite{Buen-Abad:2022kgf}, at small length scales (large wavenumber $k$), the suppression of the power spectrum compared to the model with no iDM ($f_\chi = 0$) becomes constant. When compared to the $\lcdm$ fit to $\mD$, as in Figure~\ref{fig:Pk}, the large $k$ limit is not quite a constant due to the difference in $n_s$, becoming a mild power law, but on scales ranging between $0.1$-$1 \, h/\text{Mpc}$, it is suppressed by a few percent. One can also see the dark acoustic oscillations that arise due to the iDM-DR interactions~\cite{Buen-Abad:2022kgf} in Figure~\ref{fig:Pk}. These oscillations cease once $\psi$ exits the bath at its mass threshold, but get imprinted on scales that entered the horizon prior to that ($k \sim 0.1 \, h/\text{Mpc}$). In EDE models there is an increase in power at smaller scales (exacerbating the $S_8$ problem), mostly due to a significant change in $n_s$. On the other hand, in WZDR+, as expected from similar models in which dark matter interacts with DR~\cite{Buen-Abad:2015ova,Buen-Abad:2017gxg}, the suppression becomes larger at smaller scales, decreasing with an approximately logarithmic dependence on $k$. Therefore, precise measurements of the shape of the power spectrum at scales of order the $S_8$ scale, $k\sim 10^{-1} h/\text{Mpc}$, may be able to distinguish between the \spapt and WZDR+ classes of models, as long as they can handle uncertainties due to non-linear effects. We also see from Fig.~\ref{fig:Pk} that in order to differentiate the best fit of \spapt from that of \lcdm, we would need to understand the power spectrum at these small scales at the percent level. However, since in EDE and WZDR+ the departure from \lcdm continues to grow at smaller scales, it might be possible to constrain such models using less precise probes of the power spectrum at even smaller scales, such as Lyman-$\alpha$ forests (see e.g.~\cite{Goldstein:2023gnw} for a study with EDE). In addition, all of these models affect the CMB at small scales (large $l$) in different ways, and therefore future CMB experiments, such as Simons Observatory \cite{SimonsObservatory:2019qwx}, CMB-S4 \cite{Abazajian:2019eic}, and CMB-HD \cite{Nguyen:2017zqu,Sehgal:2019nmk,Sehgal:2019ewc,Sehgal:2020yja} can play an important role in distinguishing between them.

\section{Conclusions}
\label{sec:concl}

In this work, we have performed an extensive study of a new class of interacting dark sector models, the Stepped Partially Acoustic Dark Matter (\spa) and its generalizations (\spap), by fitting them to a wide range of cosmological data. In particular, we have explored how these models compare against \lcdm when attempting to solve the tensions in measurements of $H_0$ and $S_8$. We compared fits of these models to a combination of datasets that consisted of both a baseline set of experiments and the direct measurements of $H_0$ and $S_8$ in tension with it, with fits to data that included only this baseline. We found that both \spa and \spapt are very effective in reducing the $H_0$ tension, particularly in the case where data fitted includes the direct measurements as well as the baseline dataset. The best improvements were found with \spapt, for which the best fit point to the baseline model, \ie excluding the direct $H_0$ measurement from SH0ES, already allows for $H_0$ as large as $71.4$ km/s/Mpc at 95\% CR, much larger than what is allowed within \lcdm. When we include all the data in the fit, the $\chi^2$ improvement of \spapt over \lcdm is $24.6$, which is a very significant improvement, even when taking into account the three extra parameters in the model. Ideally, the $H_0$ tension would be fully resolved in the fit to just the baseline dataset. Although this is not the case, we find it promising that even without the SH0ES prior, the tension is already reduced, and that the inclusion of the SH0ES prior, which leads to severe tension with \lcdm fits to the CMB, now allows a sizeable increase in $H_0$ without degrading the global fits of the \spa and \spapt models.

Our results also showed that the original \spa proposal was less effective at simultaneously addressing the $H_0$ and $S_8$ tensions than expected. In this model, once $\NIR$ is increased to accommodate larger $H_0$, the fraction of interacting dark matter that is allowed by the data becomes very small, and thus it cannot sufficiently lower $S_8$. In \spapt, where the step change in $\DNeff$ is smaller, the allowed range for iDM is larger which leads to a more significant reduction in $S_8$. The predicted matter power spectrum for this model differs from other proposed solutions of both tensions at small scales, and in combination with future CMB measurements, may allow one to distinguish this model from other proposals.

\hspace{1pc}

\noindent {\bf Note Added:} During the final stages of this work, Ref.~\cite{Allali:2023zbi} appeared, in which the authors study the fits to cosmological data of a phenomenological parametrization of dark sector models. One of the class of models they explore is directly related to \spa and \spap. Their analysis differs from ours in the data that is being used for the fits, the number of free parameters describing the models, and the choice of priors used for the MCMC.
In addition, shortly after v1 of this paper was posted on the arXiv, Ref.~\cite{Schoneberg:2023rnx} appeared. This paper considers an iDM-DR model with a mass threshold and with varying DR step size, both in its weak (\`{a} la WZDR+) and strongly coupled limits (\`{a} la \spa, as in this present work). In addition, the authors include ACT, SPT, and BOSS full shape datasets. Where their work overlaps with ours, the results are in excellent agreement.

\acknowledgments{

The authors thank Martin Schmaltz and Melissa Joseph for useful discussions. MBA thanks Stephanie Buen Abad for reviewing this manuscript. MBA, ZC and GMT are supported in part by the National Science Foundation under Grant Number PHY-2210361. ZC and GMT are also supported in part by the US-Israeli BSF Grant 2018236. The research of CK and TY is supported by the National Science Foundation Grant Number PHY-2210562. The authors acknowledge the \href{http://www.tacc.utexas.edu}{Texas Advanced Computing Center} (TACC) at the University of Texas at Austin for providing HPC resources that have contributed to the research results reported within this paper.
}

\appendix

\section{SPartAcous Recap}
\label{sec:apprecap}

In this section we revisit the equations governing the evolution of the iDM and DR fluids, both their background and perturbations, previously published in Ref.~\cite{Buen-Abad:2022kgf}. We rewrite them slightly in order to facilitate our derivation of the approximation schemes used in our code, and described in \App{sec:appappx}. We also briefly describe our treatment of the superhorizon initial conditions of the iDM and DR cosmological perturbations.

\subsection{Evolution Equations}
\label{subsec:appevoleqs}

\subsubsection{Background}

The iDM background equations are the same as those of CDM. The DR background equations, on the other hand, are:
\bea
    \rho_\dr & = & g_*^{\rm IR} \rho_B (T_d) \bl( 1 + r_g \hat{\rho}(x) \br) \ , \label{eq:rho_dr} \\
    P_\dr & = & \frac{1}{3} g_*^{\rm IR} \rho_B (T_d) \bl( 1 + r_g \hat{p}(x) \br) \ , \label{eq:P_dr} \\
    w_\dr \equiv \frac{P_\dr}{\rho_\dr} & = & \frac{1}{3}\frac{1 + r_g \hat{p}(x)}{1 + r_g \hat{\rho}(x)} \nonumber\\
    & = & \frac{1}{3} - \frac{r_g}{3} \frac{\hat\rho - \hat p}{1 + r_g \hat\rho} \ , \label{eq:w_dr} \\
    c_{\dr, s}^2(x) & \equiv & \frac{\dot{P}_\dr}{\dot{\rho}_\dr} = \frac{1}{3}\frac{1 + r_g \bl( \hat{p}(x) - \frac{x}{4} \hat{p}^\prime(x) \br)}{1 + r_g \bl( \hat{\rho}(x) - \frac{x}{4} \hat{\rho}^\prime(x) \br)} \nonumber\\
    & = & \frac{1}{3} - \frac{r_g}{36} \frac{x^2 \hat p}{1 + r_g \bl( \frac{3}{4}\hat\rho + \bl( \frac{1}{4} + \frac{x^2}{12} \br) \hat p \br)} \ , \label{eq:cs2}
\eea
where $r_g \equiv (g_*^{\rm UV} - g_*^{\rm IR})/g_*^{\rm IR}$, $\rho_B(T_d) \equiv \frac{\pi^2}{30} T_d^4$ is the energy density of a single bosonic degree of freedom, $x \equiv m_\psi / T_d$, and
\bea
    \hat{\rho}(x) & \equiv & \frac{x^2}{2}K_2(x) + \frac{x^3}{6}K_1(x) \ , \label{eq:rho_hat} \\
    \hat\rho^\prime & = & -\frac{x^2}{6} \bl( x K_0(x) + K_1(x) \br) \ , \\
    \hat\rho^{\prime\prime} & = & \frac{x}{6} \bl( -2 x K_0(x) + (x^2 - 1) K_1(x) \br) \ , \\
    \hat{p}(x) & \equiv & \frac{x^2}{2}K_2(x) \ , \label{eq:p_hat} \\
    \hat p^\prime & = & -\frac{x^2}{2} K_1(x) \ , \\
    \hat p^{\prime\prime} & = & \frac{x}{2} \bl( x K_0(x) - K_1(x) \br) \ .
\eea
Here $K_n(x)$ is the $n$-th order modified Bessel function of the second kind.

The evolution of the DR temperature $T_d = m_\psi/x$ can be obtained by solving for $x(a)$ from the equation that governs conservation of entropy:
\beq
    \bl( \frac{x a_t}{a} \br)^3 = 1 + \frac{r_g}{4}\bl( 3 \hat{\rho}(x) + \hat{p}(x) \br) \ . \label{eq:entropy}
\eeq
Here $a_t \equiv \frac{1}{1+z_t} \equiv T_{d0}/m_\psi$ is the scale factor at which the step begins and the $\psi$ particles start to become non-relativistic.

\subsubsection{Perturbations}

The perturbation equations for the iDM and DR fluids are given by
\bea
    \dot\delta_\idm & = & -\theta_\idm - \mM_c \ , \label{eq:dx}\\
    \dot\theta_\idm & = & -\mH \theta_\idm +\mM_e + \frac{1}{\tc} \Trx \ , \label{eq:tx}\\
    \dot\delta_\dr & = & -(1+w_\dr)(\theta_\dr + \mM_c) - 3\mH (c_{\dr, s}^2 - w_\dr)\delta_\dr \ , \label{eq:dr}\\
    \dot\theta_\dr & = & - \mH \vpi \theta_\dr + \frac{c_{\dr, s}^2}{1+w_\dr} k^2 \delta_\dr + \mM_e - \frac{1}{R \tc} \Trx \ ,\label{eq:tr}
\eea
where we have defined
\bea
    \mM_c & \equiv &
    \begin{cases}
        \dot{h}/2 & \quad \text{(synchronous gauge)}\\[2ex]
        -3 \dot\phi & \quad \text{(Newtonian gauge)}
    \end{cases}  \ , \label{eq:Mc_def} \\
    \mM_e & \equiv & 
    \begin{cases}
        0 & \quad \text{(synchronous gauge)}\\[2ex]
        k^2 \psi & \quad \text{(Newtonian gauge)}
    \end{cases}  \ , \label{eq:Me_def} \\
    R & \equiv & \frac{(1+w_\dr) \rho_\dr}{\rho_\idm} \ , \label{eq:R_def} \\
    \Trx & \equiv & \theta_\dr - \theta_\idm \ , \label{eq:Trx_def} \\
    \vpi & \equiv & (1-3 w_\dr) + \frac{\frac{d w_\dr}{d \ln a}}{1+w_\dr} = (1-3 w_\dr) + \frac{d \ln (1+w_\dr)}{d \ln a} \nonumber\\
    & = & 1 - 3 c_{\dr, s}^2 \ ,\label{eq:vpi_def}
\eea
and we have used the fact that $\frac{d}{d\tau} = \frac{d \ln a}{d \tau} \frac{d}{d \ln a} = \mH \frac{d}{d \ln a}$. The iDM--DR momentum-exchange rate and its associated (conformal) time scale are given by
\bea
    \Gamma & = & \frac{4}{3 \pi} \, \alpha_d^2 \, \ln (4/\langle \theta_{\rm min} \rangle^2) \, \frac{m_\psi^2}{m_\chi} \, x^{-2} \,e^{-x} \bl( 2 + x(2+x) \br) \ , \label{eq:Gamma} \\
    \frac{4}{\langle \theta_{\rm min} \rangle^2} & = & \frac{\pi}{g_\psi \alpha_d^3} \, \frac{K_2(x)}{2 \bl( x K_0(x) + K_1(x)  \br)^2} \ , \label{eq:theta_min} \\
    \tc & \equiv & \frac{1}{a \Gamma} \ . \label{eq:tc_def}
\eea
Note that in CLASS \cite{Blas:2011rf}, for the baryon-photon plasma, $\frac{1}{\tc^{\rm CLASS}} \equiv \frac{1}{R \tc}$.

\subsection{Initial Conditions}

In our work we consider adiabatic perturbations of the various components of the universe in the absence of curvature. These perturbations have initial superhorizon conditions that, in principle, depend on the equation of state of the fluids to which they belong. These initial conditions have been studied in detail in the past (see, for example, Refs.~\cite{Ma:1995ey,Cyr-Racine:2010qdb}). While these are straightforward to derive in the simplest case of a constant equation of state (such as for matter-like or radiation-like fluids), they are significantly more complicated in more general cases.

The DR in \spa and \spap is one such example. It undergoes an entropy dump and a step around the redshift $z_t$, which means that $w_\dr$ and $c_{\dr, s}^2$ deviate from $1/3$ around this time. One could in principle simply set the initial conditions at a redshift sufficiently before $z_t$, when $w_\dr = c_{\dr, s}^2 = 1/3$ still. However, the Hubble timescale associated with such an early time can be so small that the {\tt CLASS} code can encounter memory problems dealing with the correspondingly large number of time steps. Because of this we instead solve for the superhorizon initial conditions of the adiabatic perturbations at any time by expanding around $1/3$ to first order in the deviations $\delta w_\dr \equiv w_\dr - 1/3$ and $\delta c_{\dr, s}^2 \equiv c_{\dr, s}^2 - 1/3$. These deviations depend on $r_g$, and can reach a change relative to $1/3$ of $\lesssim -20\%$ for $r_g \leq 2$. We therefore expect that any errors introduced to the initial conditions by dropping contributions of order $\mO((\delta w_\dr)^2, \, (\delta c_{\dr, s})^2)$ and higher will be $\lesssim 4\%$.

Within this prescription the initial conditions, although taking an analytic form that can be easily implemented in {\tt CLASS}, are still rather cumbersome to write down. Therefore, we instead include a {\tt Mathematica} notebook (named {\tt spartacous\_initial\_conditions.nb} and located in the {\tt notebooks/} folder) in our {\tt class\_spartacous} code, where we systematically derive the initial conditions of the adiabatic perturbations of all the fluids present in the \spa and \spap models. These have been added to the {\tt perturbations\_initial\linebreak\_conditions} routine of the {\tt perturbations.c} module of our {\tt class\_spartacous} code.

\section{Approximation Schemes}
\label{sec:appappx}

In this appendix we describe the different regimes that the DR--iDM system experiences throughout its history, as well as the approximation schemes we employ to facilitate their numerical description within the context of the {\tt CLASS} code. Initially the DR and iDM fluids have a large momentum-exchange rate, particularly well suited to what is known as the {\it tight-coupling approximation}~\cite{Peebles:1970ag,Ma:1995ey,Blas:2011rf}. After the DR decouples from the iDM, and sufficiently deep in the matter- or dark energy-dominated era, the DR behaves as test particles in an external gravitational field, since their average energy density is negligible. This allows for another simplification of the DR evolution equations, called the {\it radiation-streaming approximation} \cite{Blas:2011rf}, which avoids significant computational efforts.

The two subsections that follow deal with the (dark) tight-coupling approximation and the (dark) radiation-streaming approximation of the DR-iDM and DR systems respectively.

\subsection{Dark Tight-Coupling Approximation (DTCA)}\label{subsec:dtca}

In the \spa model (and its \spap extension) the $\psi$ and $\chi$ particles, constituting the DR and iDM, scatter primarily via the $t$-channel exchange of the massless gauge boson $A$, under which both are charged. The relevant quantity is the momentum-exchange rate $\Gamma$ which, as long as the DR temperature $T_d$ is larger than the mass $m_\psi$, evolves in time just like the Hubble expansion rate $H$ does during radiation domination \cite{Buen-Abad:2015ova,Buen-Abad:2022kgf}. For the parameter space relevant to the \spa model $\Gamma \gg H$, which means that the DR and the iDM are tightly coupled; their ratio reaches values of the order of $\Gamma/H \sim 10^{11}$ for $\alpha_d = 10^{-3}$ and $m_\chi = 1~\TeV$.

Such a large hierarchy in the timescales relevant for the cosmological evolution of the iDM and DR components is not without its technical difficulties. Indeed, the fact that small differences in the DR and iDM velocity divergences $\theta_\dr - \theta_\idm$ are multiplied by a large rate $\Gamma$ means that the  we are in the presence of a stiff system, a kind of system which is notoriously unstable to numerical methods.

In order to address this problem it is better to think about the DR and the iDM fluids not as separate substances but as a single fluid in equilibrium. Clearly, since they are tightly coupled by the large size of $\Gamma$, any deviation from equilibrium will be quickly smoothed out over timescales $1/\Gamma \ll 1/H$. Mathematically, this means finding the equations for DR and iDM describing their mutual equilibrium, and writing their departures from the same in terms of a series expansion in the small parameter $H/\Gamma$. This method, called the {\it tight-coupling approximation}, has been successfully applied to the baryon-photon plasma, where Compton scattering tightly couples electrons and photons \cite{Ma:1995ey,Blas:2011rf}. We denote the application of this approximation scheme to our dark sector by the moniker DTCA.

In our {\tt CLASS} implementation of the \spa model, the {\tt perturbations} variable {\tt dtca\_on}/{\tt dtca\_off} controls whether the DTCA is used or not, while the {\tt precisions} variable {\tt dark\_tight\_coupling\_approximation} denotes the method by which it is implemented, currently to first-order in $H/\Gamma$. We begin our evolution of the DR--iDM system of equations with the DTCA on (unless their interactions are very small, or one or both of those fluids are not present), and turn it off when one of the following three conditions is satisfied: the ratio of Hubble to the momentum-exchange rate becomes large ($\mH/(a \Gamma) \geq 0.003$), the timescale associated with the relevant mode is also large ($k/(a \Gamma) \geq 0.0035$), or the dark temperature is close to its value at the time of $\psi$ freeze-out ($x/x_{\rm fo} \geq 0.8$, with $x_{\rm fo}$ the value of $x = m_\psi/T_d$ at freeze-out).

Throughout the rest of this section we derive the evolution equations for the perturbations in the iDM and DR fluid in the tightly-coupled limit ($\Gamma \gg H$), starting with their general versions \Eqst{eq:dx}{eq:tr}, and following the procedure laid out in Ref.~\cite{Blas:2011rf}, suitably modified to fit our case.

\subsubsection{The DTCA Equations}

The goal is to derive the equations in the DTCA, when $\tc \ll \tau, 1/k$.  Taking \Eq{eq:tx} and adding $R\,\times$\Eq{eq:tr}:
\beq\label{eq:txRtr}
    \dot\theta_\idm + R \dot\theta_\dr = - \mH \theta_\idm - \mH R \vpi \theta_\dr + R \frac{c_{\dr, s}^2}{1+w_\dr} k^2 \delta_\dr + (1+R) \mM_e \ .
\eeq
Since $\dTrx = \dot\theta_\dr - \dot\theta_\idm$, we can subtract $-R\dTrx$ from \Eq{eq:txRtr} and find:
\beq
    \dot\theta_\idm + R \dot\theta_\idm = - \mH \theta_\idm - \mH R \vpi \theta_\dr + R \frac{c_{\dr, s}^2}{1+w_\dr} k^2 \delta_\dr + (1+R) \mM_e - R \dTrx \ ,
\eeq
and thus finally arrive at:
\beq\label{eq:dtx_fin}
    \dot\theta_\idm = -\frac{1}{1+R} \bl( \mH \theta_\idm + \mH R \vpi \theta_\dr - R \frac{c_{\dr, s}^2}{1+w_\dr} k^2 \delta_\dr +  R \dTrx \br) + \mM_e \ .
\eeq

From the LHS of \Eq{eq:txRtr} we can solve for $\dot\theta_\dr$ and find:
\beq\label{eq:dtr_fin}
    \dot\theta_\dr = -\frac{1}{R} \bl( \dot\theta_\idm + \mH \theta_\idm \br) - \mH \vpi \theta_\dr + \frac{c_{\dr, s}^2}{1+w_\dr} k^2 \delta_\dr + \bl( \frac{1+R}{R} \br) \mM_e \ .
\eeq

Note that, at this stage, \Eqs{eq:dtx_fin}{eq:dtr_fin} are exact.

\subsubsection{The DTCA Slip}

We now determine the {\it slip} $\dTrx$ as a series expansion in $\tc$, and then use it in combination with \Eqs{eq:dtx_fin}{eq:dtr_fin}. Taking the combination $\tc\,\times $\Eq{eq:tr}$-\tc\,\times $\Eq{eq:tx}:
\beq
    \tc \bl( \dTrx - \mH \theta_\idm + \mH \vpi \theta_\dr - \frac{c_{\dr, s}^2}{1+w_\dr}k^2 \delta_\dr \br) + \bl( \frac{1+R}{R} \br) \Trx = 0 \ .
\eeq
Defining the useful functions
\beq
    f \equiv \frac{R}{1+R}\tc \ , \qquad g \equiv -\mH \theta_\idm + \mH \vpi \theta_\dr - \frac{c_{\dr, s}^2}{1+w_\dr} k^2 \delta_\dr \ ,
\eeq
we find that the slip satisfies the exact equation:
\beq
    f \dTrx + f g + \Trx = 0 \ .
\eeq
Note that $f/\tau$ is small; $f$ then can serve as an expansion parameter, and the idea is to solve for $\dTrx$ perturbatively in $f$. It can be shown \cite{Blas:2011rf} that this perturbative solution is:
\beq
    \Trx = \sum_n y_n  \ ; \quad y_1 = - f g \ , \quad y_{n+1} = -f \dot y_n \ ,
\eeq
\beq\label{eq:Trx_1}
    \Trx = f (-g + \dot f g + f \dot g) + \mO(f^3)
\eeq
\beq\label{eq:dTrx_1}
    \Rightarrow \dTrx = \frac{\dot f}{f} \Trx + f (- \dot g + \ddot f g + 2 \dot f \dot g + f \ddot g) + \mO(f^3) \ .
\eeq
We will only be interested in the expansion to first order, so we will then only keep $\dot g$ in the equation above. Defining the shorthand $\Delta \equiv - \frac{c_{\dr, s}^2}{1+w_\dr} k^2 \delta_\dr$, we can write:
\bea\label{eq:dg}
    \dot g & = & -\dot\mH \theta_\idm -\mH \dot\theta_\idm + \dot \mH \vpi \theta_\dr + \mH \dot \vpi \theta_\dr + \mH \vpi \dot \theta_\dr + \dot\Delta \nonumber\\
    & = & 2 \mH (1 + \vpi) \dTrx + \dot\mH \bl( -\theta_\idm + \vpi \theta_\dr \br) + \mH \dot\vpi \theta_\dr + \dot\Delta \nonumber\\
    && + \mH \bl( (1 + 2 \vpi) \bl( -\mH \theta_\idm +\mM_e + \frac{1}{\tc} \Trx \br) \br. \nonumber\\
    &&\qquad \bl. - (2 + \vpi) \bl( - \mH \vpi \theta_\dr + \frac{c_{\dr, s}^2}{1+w_\dr} k^2 \delta_\dr + \mM_e - \frac{1}{R \tc} \Trx \br) \br) \nonumber\\
    & = & \frac{\mH}{R \tc} \bl( \bl( 2+R \br) + \vpi \bl( 1+2R \br) \br) \Trx + \vpi \mH^2 \Trx + 2 \mH (1 + \vpi) \dTrx \nonumber\\
    && - \bl( \frac{\ddot a}{a} + \vpi \mH^2 \br) \bl( \theta_\idm - \vpi \theta_\dr \br) + \mH \dot\vpi \theta_\dr \nonumber\\
    && - \mH (1 - \vpi) \mM_e + \bl( \dot\Delta + \mH (2 + \vpi) \Delta \br) \ .
\eea

Noting that $\frac{\dot f}{f} = \frac{\dtc}{\tc} + \frac{\dot R/R}{1+R}$, we can combine \Eqs{eq:dg}{eq:dTrx_1}:
\bea\label{eq:dTrx_2}
    \dTrx & = & \bl( \frac{\dtc}{\tc} + \frac{\dot R/R}{1+R} - \frac{\mH \bl( \bl( 2+R \br) + \vpi \bl( 1+2R \br) \br)}{1+R} \br) \Trx \nonumber\\
    && - \frac{R \tc}{1+R} \bl[ - \bl( \frac{\ddot a}{a} + \vpi \mH^2 \br) \bl( \theta_\idm - \vpi \theta_\dr \br) - \mH (1 - \vpi) \mM_e \br. \nonumber\\
    && \quad + \mH \dot\vpi \theta_\dr + \overline\Delta \bigg] - f \bl( \vpi \mH^2 \Trx + 2 \mH (1 + \vpi) \dTrx \br) \ ,
\eea
where $\overline\Delta \equiv \dot\Delta + \mH (2 + \vpi) \Delta$. Note that the last term in \Eq{eq:dTrx_2} is of order $\mO(f^2)$ (see \Eqs{eq:Trx_1}{eq:dTrx_1}), and so we will drop it. All that remains is to compute the terms involving $\overline\Delta$, $\dot\vpi$, $\dot R$, and $\dtc$ above.

Starting with $\overline\Delta$ and using \Eqs{eq:dr}{eq:vpi_def}:
\beq
    \overline\Delta = - \mH k^2 \frac{c_{\dr, s}^2}{1+w_\dr} \delta_\dr \bl( 3 - 3 c_{\dr, s}^2 + \frac{d \ln c_{\dr, s}^2}{d \ln a} \br) + k^2 c_{\dr, s}^2 \bl( \theta_\dr + \mM_c \br) \ . \label{eq:Delta_bar_1}
\eeq

For $\dot \vpi$, using \Eq{eq:vpi_def}:
\beq
    \dot \vpi = \mH \frac{d \vpi}{d \ln a} = - 3 \mH \bl( c_{\dr, s}^2 \frac{d \ln c_{\dr, s}^2}{d \ln a} \br) \label{eq:dot_vpi_1} \ .
\eeq

For $\dot R$ we use the continuity equations $\dot \rho_\idm = -3 \mH \rho$ and $\dot \rho_\dr = -3 \mH \rho_\dr (1+w_\dr)$ to write:
\beq\label{eq:dot_R_1}
    \frac{\dot R}{R} = \mH \frac{d \ln R}{d \ln a} = - \mH \bl( 3 w_\dr - \frac{d \ln (1+w_\dr)}{d \ln a} \br) \ .
\eeq

Finally, for $\dtc$:
\beq\label{eq:dot_tc_1}
    \frac{\dtc}{\tc} = \mH \frac{d \ln \tc}{d \ln a} = - \mH \bl( \frac{d \ln \Gamma}{d \ln a} + 1 \br) \ .
\eeq

From \Eq{eq:vpi_def} and \Eqst{eq:Delta_bar_1}{eq:dot_tc_1} we know that our DTCA equations depend on $w_\dr$, $c_{\dr, s}^2$, $\Gamma$, and their derivatives with respect to the scale factor $a$. However, as shown in \Eq{eq:w_dr}, \Eq{eq:cs2}, and \Eq{eq:Gamma}, these quantities are more easily expressed in terms of analytic functions of $x = m_\psi/T_d$. Therefore, we need to convert all derivatives with respect to $a$ into derivatives with respect to $x$. This can easily be done by using the chain rule as well as \Eq{eq:entropy} in order to relate $x$ to $a$.

\subsubsection{Summary of DTCA Equations}

For convenience, we list the final form of the equations for iDM and DR here. The DTCA equations for $\theta_\idm$ and $\theta_\dr$ are:
\bea
    \dot\theta_\idm & = & -\frac{1}{1+R} \bl( \mH \theta_\idm + \mH R \vpi \theta_\dr - R \frac{c_{\dr, s}^2}{1+w_\dr} k^2 \delta_\dr +  R \dTrx \br) + \mM_e \ ,\\
    \dot\theta_\dr & = & -\frac{1}{R} \bl( \dot\theta_\idm + \mH \theta_\idm \br) - \mH \vpi \theta_\dr + \frac{c_{\dr, s}^2}{1+w_\dr} k^2 \delta_\dr + \bl( \frac{1+R}{R} \br) \mM_e \ ,
\eea
where we have found the slip $\dTrx$ to first order in $\tc/\tau$:
\bea
    \dTrx & = & \mH \bl( \frac{d \ln \tc}{d \ln a} + \frac{1}{1+R}\frac{d \ln R}{d \ln a} - \frac{\bl( 2+R \br) + \vpi \bl( 1+2R \br)}{1+R} \br) \Trx \nonumber\\
    && - \frac{R \tc}{1+R} \bl[ - \bl( \frac{\ddot a}{a} + \vpi \mH^2 \br) \bl( \theta_\idm - \vpi \theta_\dr \br) - \mH (1 - \vpi) \mM_e \br. \nonumber\\
    && \qquad + \mH^2 \frac{d \vpi}{d \ln a} \theta_\dr - \mH k^2 \frac{c_{\dr, s}^2}{1+w_\dr} \delta_\dr \bl( 3 - 3 c_{\dr, s}^2 + \frac{d \ln c_{\dr, s}^2}{d \ln a} \br) \nonumber\\
    && \qquad + k^2 c_{\dr, s}^2 \bl( \theta_\dr + \mM_c \br) \bigg] \ .
\eea

The various derivatives with respect to the scale factor $a$ are given by
\bea
    \frac{d \ln \tc}{d \ln a} & = & - \bl( \frac{d \ln \Gamma}{d \ln a} + 1 \br) \ , \\
    \frac{d \ln R}{d \ln a} & = & - \bl( 3 w_\dr - \frac{d \ln (1+w_\dr)}{d \ln a} \br) \ , \\
    \frac{d \vpi}{d \ln a} & = & - 3 \, c_{\dr, s}^2 \, \frac{d \ln c_{\dr, s}^2}{d \ln a} \ ,
\eea
which can be found from the chain rule
\beq
    \frac{d}{d \ln a} = \frac{d x}{d \ln a} \frac{d}{d x} \ ;
\eeq
used in conjunction with the following relationships between $a$ and $x$ (defining the shorthand $\alpha \equiv a/a_t$ and dropping the argument $x$ in $K_n(x)$):
\bea
    \alpha^3 & \equiv & \bl( \frac{a}{a_t} \br)^3 = \frac{x^3}{1 + \frac{r_g}{4}\bl( 3 \hat{\rho} + \hat{p} \br)} \ , \\
    \frac{d x}{d \ln a} & = & \frac{x}{1 + \frac{r_g}{24} \alpha^3 x K_2} \ ,
\eea
and the following derivatives with respect to $x$:
\bea
    \frac{d \ln (1+w_\dr)}{d x} & = & \frac{w_\dr}{1+w_\dr} \bl( \frac{r_g \hat p^\prime}{1 + r_g \hat p} - \frac{r_g \hat \rho^\prime}{1 + r_g \hat \rho} \br) \ , \\
    \frac{d \ln c_{\dr, s}^2}{d x} & = & \frac{r_g \bl( \frac{3}{4}\hat{p}^\prime - \frac{x}{4} \hat p^{\prime\prime} \br)}{1 + r_g \bl( \hat{p} - \frac{x}{4} \hat{p}^\prime \br)} - \frac{r_g \bl( \frac{3}{4}\hat{\rho}^\prime - \frac{x}{4} \hat\rho^{\prime\prime} \br)}{1 + r_g \bl( \hat{\rho} - \frac{x}{4} \hat{\rho}^\prime \br)} \ , \\
    \frac{d \ln \Gamma}{d x} & = & - 1 - \frac{2}{x} + \frac{2(1+x)}{2+x(2+x)} + \frac{1}{\ln \bl( \frac{4}{\langle \theta_{\rm min} \rangle^2} \br)} \frac{d \ln (\frac{4}{\langle \theta_{\rm min} \rangle^2}) }{d x} \ , \\
    \frac{d \ln (\frac{4}{\langle \theta_{\rm min} \rangle^2}) }{d x} & = & \frac{-3 K_0^2 + \bl(\frac{x^2 - 8}{x} \br) K_0 K_1 + \bl( \frac{3 x^2 - 4}{x^2} \br) K_1^2 + K_2^2}{K_2 \bl( x K_0 + K_1 \br)} \ .
\eea

\subsection{Dark Radiation Streaming Approximation (DRSA)}\label{subsec:drsa}

Soon after matter-radiation equality, the energy density in the relativistic particles, including DR, becomes negligible. Because of this, once DR is decoupled from the iDM it effectively behaves as a fluid in an external gravitational field. As it turns out, finding the exact behavior of the DR perturbations in this regime is both unimportant and computationally prohibitive. It is unimportant because DR is made up of dark, unobservable particles that lead to no observable signature of their own ({\it e.g.} a dark CMB), and it is computationally expensive due to the fast oscillations (hard to calculate precisely) taking place in its perturbations, which arise due to the large value of $k \tau$. In this limit one can find the non-oscillating component of the DR equations using a {\it dark radiation-streaming approximation}, which we abbreviate as DRSA, and which is based on the methods described in Ref.~\cite{Blas:2011rf}.\footnote{Note that, despite the name, this approximation also works for our self-interacting DR. Indeed, in deriving the non-oscillatory part of the solution to the equations of motion of the DR, the assumptions about the DR are that ({\it i.}) its energy density is subdominant (trivially true well after matter-radiation equality), ({\it ii.}) its shear is negligible (always true of self-interacting fluids), and ({\it iii.}) its equation of state satisfies $w_\dr = c_{\dr,s}^2 = 1/3$ (still valid at times well past the step).}

In our modified version of {\tt CLASS} including the \spa model, the {\tt perturbations} variable {\tt drsa\_on}/{\tt drsa\_off} denotes whether the DRSA is turned on or off, while the\linebreak {\tt precisions} variable {\tt dark\_radiation\_streaming\_approximation} refers to the method of its implementation, currently a suitably modified version of the approximation employed for the neutrinos in Ref.~\cite{Blas:2011rf} and the pre-existing {\tt idr} (interacting DR) fluid in the newest version of {\tt CLASS}. In our code, the DRSA is only turned on when all of the following conditions are true: the relevant mode is deep inside the horizon ($k \tau > 44.0$), the universe has evolved sufficiently past matter-radiation equality ($\tau/\tau_\eq > 6.0$), the redshift is significantly past $z_\tr$, when the step in the DR took place ($\tau/\tau(z = z_\tr/30) > 5.0$), and the DTCA is turned off.

\section{Numerical Results}
\label{sec:appnum}

\begin{figure}[tbh!]
    \advance\leftskip-4pc
	\includegraphics[width=1.25\linewidth]{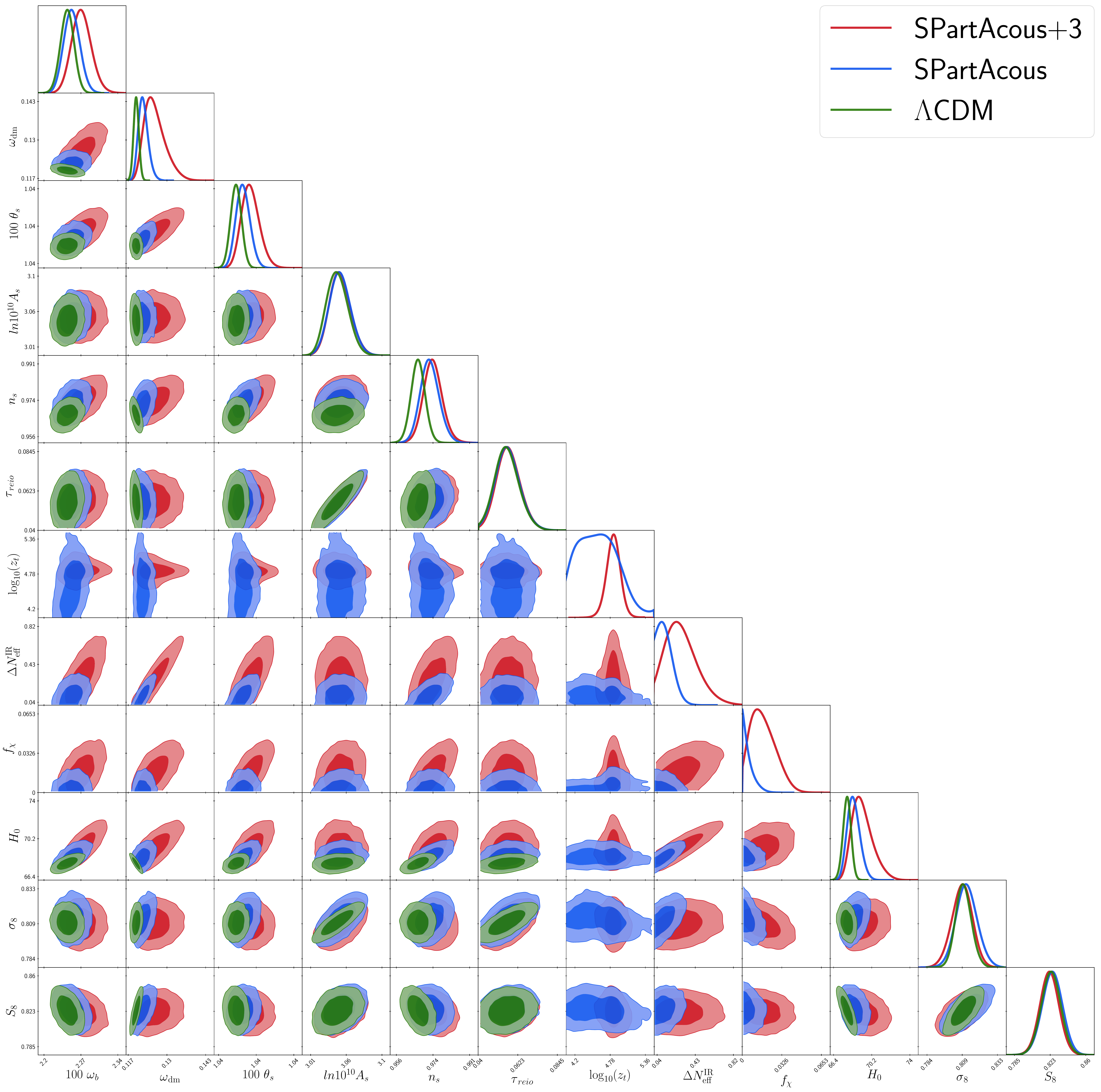}
	\caption{1D and 2D posterior distributions of the parameters of the \lcdm ({\bf green}), \spa ({\bf blue}), and \spapt ({\bf red}) models, fitted to the $\mD$ dataset.}
	\label{fig:D-triangle}
\end{figure}

\begin{figure}[tbh!]
    \advance\leftskip-4pc
	\includegraphics[width=1.25\linewidth]{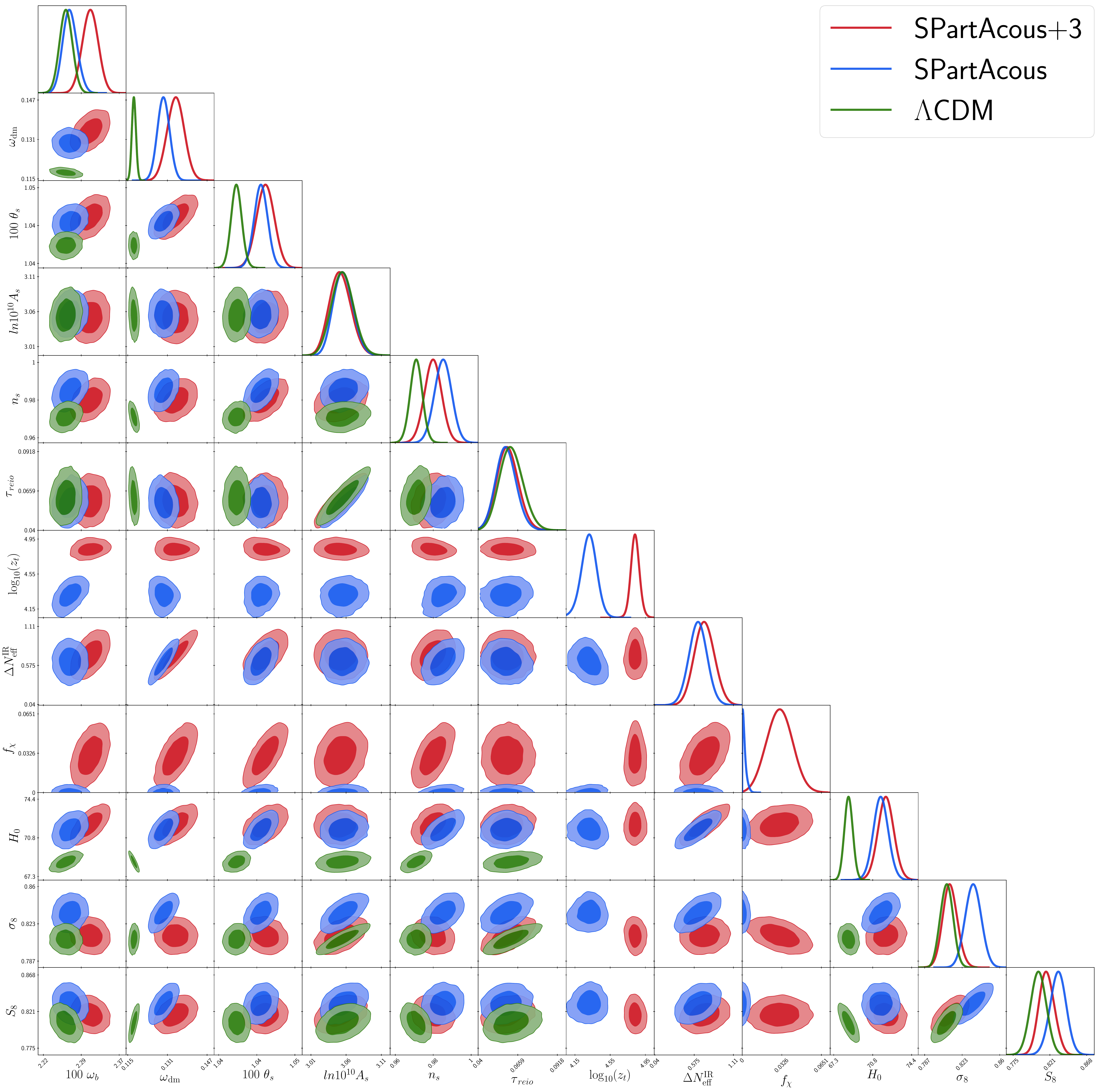}
	\caption{1D and 2D posterior distributions of the parameters of the \lcdm ({\bf green}), \spa ({\bf blue}), and \spapt ({\bf red}) models, fitted to the $\mDH$ dataset.}
	\label{fig:DH-triangle}
\end{figure}

\begin{figure}[tbh!]
    \advance\leftskip-4pc
	\includegraphics[width=1.25\linewidth]{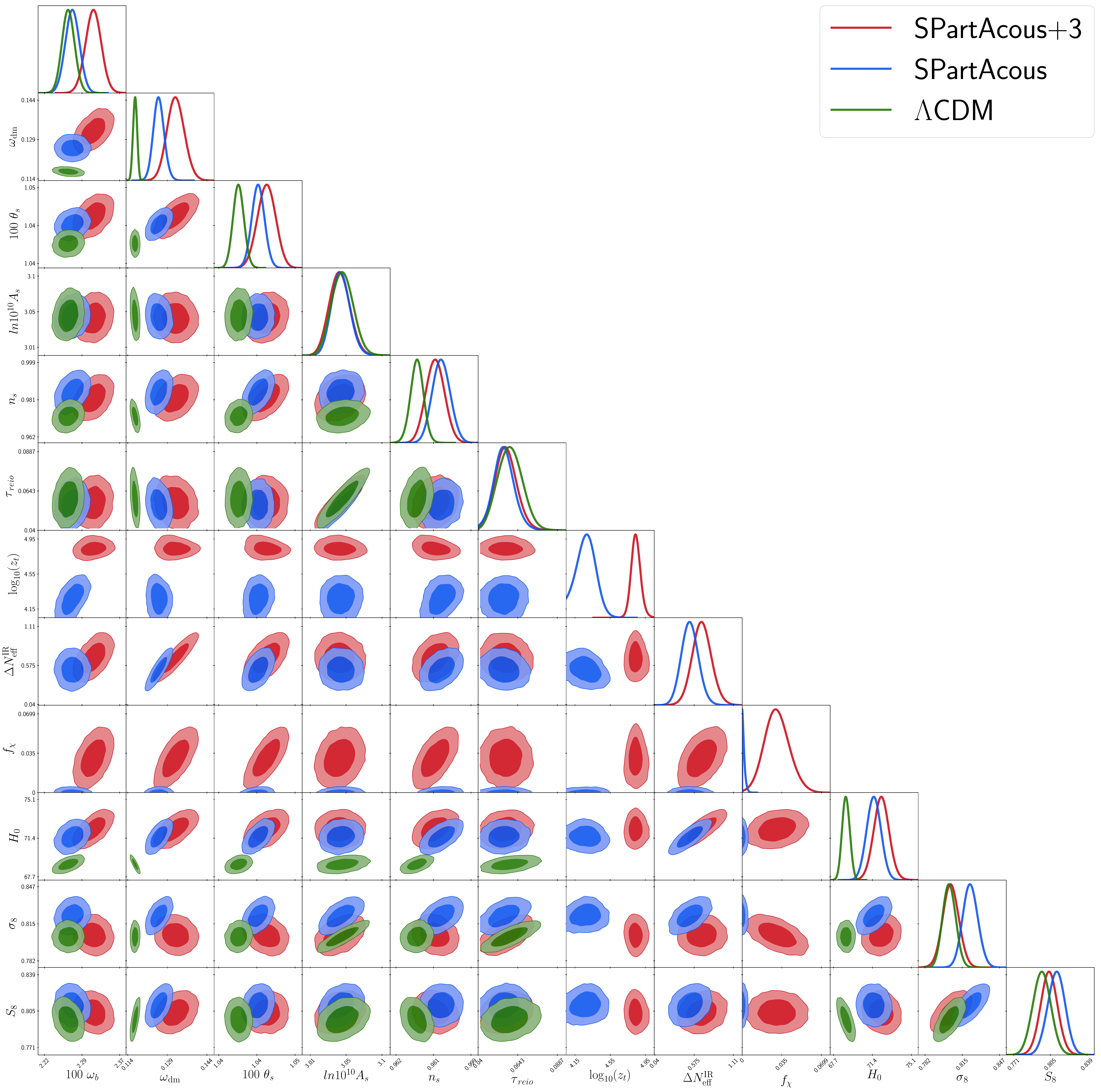}
	\caption{1D and 2D posterior distributions of the parameters of the \lcdm ({\bf green}), \spa ({\bf blue}), and \spapt ({\bf red}) models, fitted to the $\mDHS$ dataset.}
	\label{fig:DHS-triangle}
\end{figure}

In this section we provide comprehensive triangle plots and parameter values obtained from our MCMC numerical results. We illustrate 1D and 2D posterior distributions for the $\mD$, $\mDH$ and $\mDHS$ dataset in Figures~\ref{fig:D-triangle}--\ref{fig:DHS-triangle} respectively. The corresponding best-fit, mean, and $\pm 1\sigma$ values for the fits of these datasets are summarized in Tables~\ref{tab:bf_ddhs}, \ref{tab:ms_d} and \ref{tab:ms_dhs}. Table~\ref{tab:val_spar} contain the best-fit, mean, and $\pm 1\sigma$ values of the \spa model applied to all the datasets ($\mD$, $\mDH$, $\mDHS$). For the \spapt model, please refer to Table~\ref{tab:val_spar+3} for the respective values.
Throughout this study, we make the assumption of neglecting the additional relativistic degrees of freedom at the time of BBN, thereby removing their effect on the Standard Model abundance of primordial helium.

\begin{table}[h!]
\centering
\begin{adjustbox}{max width=\columnwidth}
\begin{tabular}{|c|cc|cc|cc|}
\toprule\toprule
Dataset                       & \multicolumn{2}{c|}{$\mD$}                                               & \multicolumn{2}{c|}{$\mDH$}                                      & \multicolumn{2}{c|}{$\mDHS$}                                     \\ \midrule
Value                         & \multicolumn{1}{c|}{Mean$\pm1\sigma$}                        & Best-fit  & \multicolumn{1}{c|}{Mean$\pm1\sigma$}                & Best-fit  & \multicolumn{1}{c|}{Mean$\pm1\sigma$}                & Best-fit  \\ \midrule
$100~\theta_s$                 & \multicolumn{1}{c|}{$1.0423_{-0.0004}^{+0.0003}$}            & $1.0422$  & \multicolumn{1}{c|}{$1.0434_{-0.0004}^{+0.0004}$}    & $1.0433$  & \multicolumn{1}{c|}{$1.0431_{-0.0003}^{+0.0003}$}    & $1.0431$  \\
$100~\omega_{b}$              & \multicolumn{1}{c|}{$2.252_{-0.015}^{+0.015}$}               & $2.251$   & \multicolumn{1}{c|}{$2.270_{-0.015}^{+0.014}$}       & $2.272$   & \multicolumn{1}{c|}{$2.275_{-0.015}^{+0.014}$}       & $2.277$   \\
$\omega_\dm$                  & \multicolumn{1}{c|}{$0.1216_{-0.0020}^{+0.0014}$}            & $0.1213$  & \multicolumn{1}{c|}{$0.1293_{-0.0024}^{+0.0025}$}    & $0.1292$  & \multicolumn{1}{c|}{$0.1259_{-0.0022}^{+0.0020}$}    & $0.1255$  \\
$\ln 10^{10} A_s$             & \multicolumn{1}{c|}{$3.050_{-0.015}^{+0.013}$}               & $3.048$   & \multicolumn{1}{c|}{$3.056_{-0.014}^{+0.013}$}       & $3.054$   & \multicolumn{1}{c|}{$3.047_{-0.013}^{+0.012}$}       & $3.046$   \\
$n_s$                         & \multicolumn{1}{c|}{$0.9722_{-0.0046}^{+0.0042}$}            & $0.9720$  & \multicolumn{1}{c|}{$0.9857_{-0.0044}^{+0.0045}$}    & $0.9859$  & \multicolumn{1}{c|}{$0.9848_{-0.0043}^{+0.0044}$}    & $0.9848$  \\
$\tau_\mathrm{reio}$          & \multicolumn{1}{c|}{$0.05668_{-0.00720}^{+0.00631}$}         & $0.05583$ & \multicolumn{1}{c|}{$0.05832_{-0.00734}^{+0.00602}$} & $0.05734$ & \multicolumn{1}{c|}{$0.05594_{-0.00664}^{+0.00599}$} & $0.05585$ \\
$\NIR$                        & \multicolumn{1}{c|}{$0.1205_{-0.1194}^{+0.0331}$}            & $0.1210$  & \multicolumn{1}{c|}{$0.6287_{-0.1267}^{+0.1276}$}    & $0.6384$  & \multicolumn{1}{c|}{$0.5182_{-0.1212}^{+0.1136}$}    & $0.5083$  \\
$f_\chi [\%]$                 & \multicolumn{1}{c|}{$0.5216_{-0.5216}^{+0.1090}$}            & $0.0071$  & \multicolumn{1}{c|}{$0.1885_{-0.1884}^{+0.0428}$}    & $0.0011$  & \multicolumn{1}{c|}{$0.1453_{-0.1453}^{+0.0317}$}    & $0.0015$   \\
$\log_{10}(z_t)$              & \multicolumn{1}{c|}{$4.581_{-\mathrm{nan}}^{+\mathrm{nan}}$} & $4.360$   & \multicolumn{1}{c|}{$4.309_{-0.082}^{+0.094}$}       & $4.292$   & \multicolumn{1}{c|}{$4.268_{-0.110}^{+0.120}$}       & $4.263$   \\ \midrule
$M_B$                         & \multicolumn{1}{c|}{$-19.394_{-0.021}^{+0.016}$}             & $-19.392$ & \multicolumn{1}{c|}{$-19.301_{-0.019}^{+0.020}$}     & $-19.299$ & \multicolumn{1}{c|}{$-19.305_{-0.019}^{+0.019}$}     & $-19.305$ \\
$H_0~[\km/\seg/\Mpc]$              & \multicolumn{1}{c|}{$68.42_{-0.70}^{+0.56}$}                 & $68.46$   & \multicolumn{1}{c|}{$71.56_{-0.70}^{+0.69}$}         & $71.66$   & \multicolumn{1}{c|}{$71.54_{-0.67}^{+0.66}$}         & $71.55$   \\
$\sigma_8$                    & \multicolumn{1}{c|}{$0.8122_{-0.0072}^{+0.0075}$}            & $0.8156$  & \multicolumn{1}{c|}{$0.8342_{-0.0077}^{+0.0076}$}    & $0.8354$  & \multicolumn{1}{c|}{$0.8218_{-0.0065}^{+0.0065}$}    & $0.8228$  \\
$S_8$                         & \multicolumn{1}{c|}{$0.8247_{-0.0106}^{+0.0104}$}            & $0.8266$  & \multicolumn{1}{c|}{$0.8315_{-0.0102}^{+0.0103}$}    & $0.8314$  & \multicolumn{1}{c|}{$0.8103_{-0.0081}^{+0.0079}$}    & $0.8103$  \\ \midrule\midrule
$\chi^2_\mathrm{CMB}$         & \multicolumn{2}{c|}{$2765.06$}                                           & \multicolumn{2}{c|}{$2768.41$}                                   & \multicolumn{2}{c|}{$2768.68$}                                   \\
$\chi^2_\mathrm{Pantheon}$    & \multicolumn{2}{c|}{$1025.94$}                                           & \multicolumn{2}{c|}{$1025.69$}                                   & \multicolumn{2}{c|}{$1025.85$}                                   \\
$\chi^2_\mathrm{BAO}$         & \multicolumn{2}{c|}{$5.30$}                                              & \multicolumn{2}{c|}{$6.01$}                                      & \multicolumn{2}{c|}{$7.88$}                                      \\
$\chi^2_\mathrm{Pl. lensing}$ & \multicolumn{2}{c|}{$9.09$}                                              & \multicolumn{2}{c|}{$10.33$}                                     & \multicolumn{2}{c|}{$10.86$}                                     \\
$\chi^2_{S_8}$                & \multicolumn{2}{c|}{$-$}                                                 & \multicolumn{2}{c|}{$-$}                                         & \multicolumn{2}{c|}{$6.75$}                                      \\
$\chi^2_\mathrm{SH0ES}$       & \multicolumn{2}{c|}{$-$}                                                 & \multicolumn{2}{c|}{$2.89$}                                      & \multicolumn{2}{c|}{$3.68$}                                      \\ \midrule
$\chi^2_\mathrm{tot}$         & \multicolumn{2}{c|}{$3805.39$}                                           & \multicolumn{2}{c|}{$3813.33$}                                   & \multicolumn{2}{c|}{$3823.70$}                                   \\ \bottomrule\bottomrule
\end{tabular}
\end{adjustbox}
\caption{Mean$\pm 1\sigma$ and best-fit values of the \spa model to all the datasets studied in this paper. Note that the $\pm 1 \sigma$ values of the $\log_{10} (z_t)$ parameter in \spa are ``nan''. This is because of the very non-Gaussian nature of its posterior.}
\label{tab:val_spar}
\end{table}

\begin{table}[h!]
\centering
\begin{adjustbox}{max width=\columnwidth}
\begin{tabular}{|c|cc|cc|cc|}
\toprule\toprule
Dataset                       & \multicolumn{2}{c|}{$\mD$}                                       & \multicolumn{2}{c|}{$\mDH$}                                      & \multicolumn{2}{c|}{$\mDHS$}                                     \\ \midrule
Value                         & \multicolumn{1}{c|}{Mean$\pm1\sigma$}                & Best-fit  & \multicolumn{1}{c|}{Mean$\pm1\sigma$}                & Best-fit  & \multicolumn{1}{c|}{Mean$\pm1\sigma$}                & Best-fit  \\ \midrule
$100~\theta_s$                 & \multicolumn{1}{c|}{$1.0427_{-0.0005}^{+0.0004}$}    & $1.0427$  & \multicolumn{1}{c|}{$1.0436_{-0.0005}^{+0.0004}$}    & $1.0436$  & \multicolumn{1}{c|}{$1.0436_{-0.0005}^{+0.0005}$}    & $1.0436$  \\
$100~\omega_{b}$              & \multicolumn{1}{c|}{$2.271_{-0.021}^{+0.018}$}       & $2.268$   & \multicolumn{1}{c|}{$2.311_{-0.016}^{+0.016}$}       & $2.310$   & \multicolumn{1}{c|}{$2.316_{-0.016}^{+0.016}$}       & $2.317$   \\
$\omega_\dm$                  & \multicolumn{1}{c|}{$0.1257_{-0.0043}^{+0.0028}$}    & $0.1250$  & \multicolumn{1}{c|}{$0.1341_{-0.0035}^{+0.0034}$}    & $0.1342$   & \multicolumn{1}{c|}{$0.1323_{-0.0035}^{+0.0032}$}    & $0.1327$  \\
$\ln 10^{10} A_s$             & \multicolumn{1}{c|}{$3.051_{-0.015}^{+0.013}$}       & $3.050$   & \multicolumn{1}{c|}{$3.052_{-0.015}^{+0.014}$}       & $3.052$   & \multicolumn{1}{c|}{$3.045_{-0.014}^{+0.013}$}       & $3.045$   \\
$n_s$                         & \multicolumn{1}{c|}{$0.9735_{-0.0051}^{+0.0044}$}    & $0.9737$  & \multicolumn{1}{c|}{$0.9804_{-0.0046}^{+0.0046}$}    & $0.9804$  & \multicolumn{1}{c|}{$0.9817_{-0.0047}^{+0.0046}$}    & $0.9822$  \\
$\tau_\mathrm{reio}$          & \multicolumn{1}{c|}{$0.05713_{-0.00688}^{+0.00644}$} & $0.05655$ & \multicolumn{1}{c|}{$0.05928_{-0.00734}^{+0.00674}$} & $0.05864$ & \multicolumn{1}{c|}{$0.05715_{-0.00709}^{+0.00626}$} & $0.05674$ \\
$\NIR$                        & \multicolumn{1}{c|}{$0.2892_{-0.2035}^{+0.1156}$}    & $0.2486$  & \multicolumn{1}{c|}{$0.7165_{-0.1378}^{+0.1343}$}    & $0.7158$  & \multicolumn{1}{c|}{$0.6802_{-0.1418}^{+0.1298}$}    & $0.6908$  \\
$f_\chi [\%]$                 & \multicolumn{1}{c|}{$1.7477_{-1.2927}^{+0.7382}$}    & $1.6918$   & \multicolumn{1}{c|}{$3.0751_{-1.0778}^{+1.0583}$}    & $3.1660$   & \multicolumn{1}{c|}{$3.0854_{-1.1224}^{+1.0788}$}    & $3.2616$   \\
$\log_{10}(z_t)$              & \multicolumn{1}{c|}{$4.825_{-0.094}^{+0.085}$}       & $4.814$   & \multicolumn{1}{c|}{$4.834_{-0.049}^{+0.046}$}       & $4.833$   & \multicolumn{1}{c|}{$4.841_{-0.056}^{+0.046}$}       & $4.840$   \\ \midrule
$M_B$                         & \multicolumn{1}{c|}{$-19.365_{-0.037}^{+0.026}$}     & $-19.371$ & \multicolumn{1}{c|}{$-19.29_{-0.022}^{+0.022}$}      & $-19.285$ & \multicolumn{1}{c|}{$-19.280_{-0.022}^{+0.022}$}     & $-19.279$ \\
$H_0~[\km/\seg/\Mpc]$              & \multicolumn{1}{c|}{$69.34_{-1.20}^{+0.84}$}         & $69.07$   & \multicolumn{1}{c|}{$72.01_{-0.76}^{+0.74}$}         & $71.98$   & \multicolumn{1}{c|}{$72.25_{-0.76}^{+0.75}$}         & $72.26$   \\
$\sigma_8$                    & \multicolumn{1}{c|}{$0.8086_{-0.0073}^{+0.0073}$}    & $0.8080$  & \multicolumn{1}{c|}{$0.8116_{-0.0075}^{+0.0074}$}    & $0.8108$  & \multicolumn{1}{c|}{$0.8047_{-0.0071}^{+0.0069}$}    & $0.8039$  \\
$S_8$                         & \multicolumn{1}{c|}{$0.8219_{-0.0100}^{+0.0102}$}    & $0.8224$  & \multicolumn{1}{c|}{$0.8175_{-0.0098}^{+0.0097}$}    & $0.8172$  & \multicolumn{1}{c|}{$0.8035_{-0.0083}^{+0.0080}$}    & $0.8036$  \\ \midrule\midrule
$\chi^2_\mathrm{CMB}$         & \multicolumn{2}{c|}{$2764.02$}                                   & \multicolumn{2}{c|}{$2768.11$}                                   & \multicolumn{2}{c|}{$2769.39$}                                   \\
$\chi^2_\mathrm{Pantheon}$    & \multicolumn{2}{c|}{$1026.38$}                                   & \multicolumn{2}{c|}{$1025.81$}                                   & \multicolumn{2}{c|}{$1025.78$}                                   \\
$\chi^2_\mathrm{BAO}$         & \multicolumn{2}{c|}{$5.61$}                                      & \multicolumn{2}{c|}{$5.13$}                                      & \multicolumn{2}{c|}{$5.47$}                                      \\
$\chi^2_\mathrm{Pl. lensing}$ & \multicolumn{2}{c|}{$8.89$}                                      & \multicolumn{2}{c|}{$9.19$}                                      & \multicolumn{2}{c|}{$10.03$}                                     \\
$\chi^2_{S_8}$                & \multicolumn{2}{c|}{$-$}                                         & \multicolumn{2}{c|}{$-$}                                         & \multicolumn{2}{c|}{$4.74$}                                      \\
$\chi^2_\mathrm{SH0ES}$       & \multicolumn{2}{c|}{$-$}                                         & \multicolumn{2}{c|}{$1.43$}                                      & \multicolumn{2}{c|}{$0.93$}                                      \\ \midrule
$\chi^2_\mathrm{tot}$         & \multicolumn{2}{c|}{$3804.90$}                                   & \multicolumn{2}{c|}{$3809.68$}                                   & \multicolumn{2}{c|}{$3816.34$}                                   \\ \bottomrule\bottomrule
\end{tabular}
\end{adjustbox}
\caption{Mean$\pm 1\sigma$ and best-fit values of the \spapt model to all the datasets studied in this paper.}
\label{tab:val_spar+3}
\end{table}

\clearpage

\bibliographystyle{JHEP}
\bibliography{spartacous-bib.bib}

\end{document}